\documentclass[structabstract]{aa}  
\usepackage{graphicx}
\usepackage{txfonts}
\usepackage{epstopdf}

\usepackage{longtable}
\usepackage{lscape}
\usepackage{natbib}

\usepackage{color}
\def\Msun{\hbox{$\thinspace M_{\odot}$}}
\def\Rsun{\hbox{$\thinspace R_{\odot}$}}
\def\Teff{\hbox{$\thinspace T_{\mathrm{eff}}$}}
\def\kms{\hbox{$\thinspace {\mathrm{km~s^{-1}}}$}}
\def\ms{\hbox{$\thinspace {\mathrm{m~s^{-1}}}$}}
\def\ALi{\hbox{$\thinspace A (\mathrm{Li})$}}
\def\vmic{\hbox{$\thinspace v_{\mathrm{mic}}$}}
\def\zetahm{\hbox{$\thinspace \zeta_{\mathrm{HM}}$}}

\bibpunct{(}{)}{;}{a}{}{,} % to follow the A&A style
\begin{document}

\title{The Penn State - Toru\'n Centre for Astronomy Planet Search stars. 
\thanks{Based on observations obtained with the Hobby-Eberly Telescope, 
which is a joint project of the University of Texas at Austin, the Pennsylvania State University, 
Stanford University, Ludwig-Maximilians-Universit\"at M\"unchen, and Georg-August-Universit\"at G\"ottingen.}
\thanks{
Based on observations made with the Italian Telescopio Nazionale Galileo (TNG) operated on the island of La Palma 
by the Fundaci\'on Galileo Galilei of the INAF (Istituto Nazionale di Astrofisica) at the Spanish Observatorio del Roque 
de los Muchachos of the Instituto de Astrof\'{\i}sica de Canarias.
}
\thanks{
Based on observations made with the Nordic Optical Telescope, operated
on the island of La Palma jointly by Denmark, Finland, Iceland,
Norway, and Sweden, in the Spanish Observatorio del Roque de los
Muchachos of the Instituto de Astrof\'{\i}sica de Canarias.
}
\thanks{
Based on observations obtained with the HERMES spectrograph, which is supported by the Fund 
for Scientific Research of Flanders (FWO), Belgium, the Research Council of K.U.Leuven, Belgium, 
the Fonds National Recherches Scientific (FNRS), Belgium, the Royal Observatory of Belgium, 
the Observatoire de Gen\'{e}ve, Switzerland and the Th\"uringer Landessternwarte Tautenburg, Germany.}
}

   \subtitle{II. Lithium abundance analysis of the Red Giant Clump sample} 

   \author{M. Adam\'ow
          \inst{1}
          \and
	   A. Niedzielski
          \inst{1}
          	              \and
          E. Villaver
          \inst{2}
                \and
           A. Wolszczan
	  \inst{3,4}
          	              \and
          G. Nowak
          \inst{1,5,6}
                   }

   \institute{Toru\'n Centre for Astronomy, Faculty of Physics, Astronomy and Informatics, Nicolaus Copernicus University, Grudziadzka 5, 87-100 Toru\'n, Poland.
              \email{adamow@astri.umk.pl, aniedzi@astri.umk.pl, grzenow@astri.umk.pl}
                                    \and
         Departamento de F\'{\i}sica Te\'orica, Universidad Aut\'onoma de Madrid, Cantoblanco 28049 Madrid, Spain.
         \email{Eva.Villaver@uam.es}
         \and
             Department of Astronomy and Astrophysics, Pennsylvania State University, 525 Davey Laboratory, University Park, PA 16802, USA.
          \email{alex@astro.psu.edu}
	 \and
             Center for Exoplanets and Habitable Worlds, Pennsylvania State University, 525 Davey Laboratory, University Park, PA 16802, USA.
             \and
              Instituto de Astrof\'isica de Canarias, C/ v\'ia L\'actea, s/n, E-38205 La Laguna, Tenerife, Spain.
              \and
              Departamento de Astrof\'isica, Universidad de La Laguna, Av. Astrof\'isico Francisco S\'anchez, s/n, E-38206 La Laguna, Tenerife, Spain
             }

 \date{Received ; accepted }

% \abstract{}{}{}{}{} 
% 5 {} token are mandatory
 
  \abstract
  % context heading (optional), leave it empty if necessary  
{Standard stellar evolution theory does not predict existence of Li-rich giant stars. Several mechanisms for Li-enrichment have been 
proposed to operate at certain locations inside some stars. The actual mechanism operating in real stars is still unknown.}
{Using the sample of 348 stars from the  Penn State - Toru\'n Centre for Astronomy Planet Search, for which uniformly
 determined atmospheric parameters are available, with  chemical abundances and
rotational velocities presented here, we investigate various channels of Li enrichment in giants. 
 We also study Li-overabundant giants in more detail in search for origin of their peculiarities.  }
{Our work is based on the Hobby-Eberly Telescope spectra obtained with the High Resolution Spectrograph, 
which we use for determination 
of abundances and rotational velocities. The Li abundance was determined from the $^7$Li $\lambda$670.8~nm line, 
while we use a more extended set of lines for $\alpha$-elements abundances.
In a series of Kolmogorov-Smirnov tests, we compare Li-overabundant giants with other stars in the sample.
We also  use available IR photometric  and kinematical data in search for evidence of mass-loss.
We investigate properties of the most Li-abundant giants in more detail by using multi-epoch precise radial velocities.}
{We present Li and $\alpha$-elements abundances, as well as rotational velocities for 348 stars. We  detected Li in 92 stars, 
of which 82 are giants. Eleven of them show significant Li abundance $\ALi_{NLTE}>1.4$ and  seven
of them are Li-overabundant objects, according to common criterion of $\ALi>1.5$ and their location on HR diagram,
including TYC 0684-00553-1 and TYC 3105-00152-1, which are two giants with Li abundances close to meteoritic level.
For another 271 stars, upper limits of Li abundance are presented. 
We confirmed three objects with increased stellar rotation. 
We show that Li-overabundant giants are among the most massive stars from our sample and show larger than 
average effective temperatures. They are indistinguishable from the complete sample in terms of their distribution 
of luminosity, metallicity,  rotational velocities, and $\alpha$-elements abundances.
Our results do not point out to one specific Li-enrichment mechanism 
operating in our sample of giants. 
On the contrary, in some cases, we cannot identify fingerprints of any of known scenarios.
We show, however, that the four most Li-rich giants
in our sample either have low-mass companions or have radial velocity variations at the level of $\sim100 \ms$, which 
strongly suggests that the presence of companions is an important factor in the Li-enrichment processes in giants. 
}
{}


   \maketitle
%
%________________________________________________________________

\section{Introduction}

Lithium, one of the three elements synthesized during the Big Bang Nucleosynthesis, is easily destroyed 
in the stellar interiors at temperatures  exceeding $\approx 2.5 \times 10^6 K$.
Abundance of $\ALi$\footnote{$\ALi=\log \frac{n(\mathrm{Li})}{n(\mathrm{H})}+12$} = 3.3~dex,
which represents the value found in the Solar System meteorites \citep{Asplund2009}, is considered 
the reference limit for population I stars. 

The abundance of lithium, which is in the formation of a star, is expected to be preserved in the outermost surface layers of main sequence (MS) stars.
As soon as the star evolves into the red giant branch (RGB) phase,  the deepening of the convective zone brings
 material that has been exposed to  high temperatures in the stellar interior  to the stellar surface.
Products of H-burning enrich the external layers with nitrogen and helium and the abundance of lithium 
is expected to drop to $\ALi=1.5$~dex, assuming that the star left the MS with a Li abundance close to 3.3~dex \citep{Iben1967}.
However, most observations of FGK dwarf stars reveal  lithium abundances below the meteoritic reference value (e.g. \citealt{Ghezzi2010, Ramirez2012}), which indicates the 
existence of non-standard mixing mechanisms operating during the MS phase.
This is probably the most important reason why the overwhelming majority of giant stars present 
lithium abundances below the 1.5~dex value predicted by the standard evolution theory \citep{Iben1967}. 

There are not many processes by which lithium can be produced in stars. 
However, 1-2 \% of all observed giants (see e.g. \citealt{Kumar2011,Lebzelter2012} and references therein) 
are lithium rich (stars with $\ALi>1.5$~dex). The abundance of lithium in such Li-rich giant stars 
can even exceed the amount of lithium found in meteorites.

The Cameron-Fowler (CF) mechanism \citep{CamFow1971} is usually invoked to explain the lithium overabundance in giant stars. 
Lithium production through this process involves, firstly,  transport of $^3$He from the outer layers to regions that are hot enough 
to ignite the $^3$He$(\alpha,\gamma)^7$Be burning reaction, and, secondly,  transport of beryllium to the outer layers, 
where it decays to lithium, according to the reaction $^7$Be($e^-, \nu$)$^7$Li.  
The CF mechanism is known to operate in stars with masses $4-8\Msun$ \citep{CharBal2000} 
during their evolution on the asymptotic giant branch (AGB) in which deeper convection reaches
the layers of nuclear burning (hot bottom burning).

The  lower mass  giants lack 
an additional mixing mechanism rather than convection itself to initiate the CF process, 
since the convection zone is too shallow there to feed $^3$He to the hotter stellar interior. 

Extra-mixing is expected to occur at a specific location on the Hertzsprung-Russell (HR) diagram, the luminosity function bump (LFB), 
which is associated with the removal of the molecular discontinuity that stems from the dredge-up processes.
\cite{CharBal2000} and \cite{Kumar2011} have claimed that Li-rich giants concentrate 
in a narrow luminosity range associated with the LFB on the red giant branch and/or the red giant clump (RGC) region,
\footnote{Formed by stars in the stage of central helium burning \citep{Cannon1970}.} 
while other authors show that Li-rich giants can be found almost anywhere along the RGB
[see e.g. \cite{Anthony-Twarog2013, Martell2013, Lebzelter2012}]. 
The Li enhancement mechanism on the RGB is thus still not clear and other processes explored to explain extra-mixing 
in RGB stars do not seem to be efficient enough. Thermohaline convection \citep{CharZahn2007} seems too slow 
to be responsible for Li overabundance in giants, and magneto--thermohaline mixing \citep{DenPinMac2009},
and magnetic buoyancy \citep{Busso2007} cannot explain observed values when $\ALi >2.5$~dex.

External pollution offers an alternative to the CF production of lithium by a mechanism in which
the chemical composition of the stellar photosphere might be altered through accretion of external
material from type II supernova (SN) explosions \citep{WoosleyWeaver1995}, planets \citep{Alexander1967, SiessLivio1999}, 
or from a more evolved stellar companion \citep{SacBoo1999}, such as an AGB star producing Li via CF mechanism. 
Each of these processes should present additional chemical features, besides the Li abundances, 
associated with them, that is, type II supernovae pollution should show increased levels of $\alpha$-elements 
(see e.g. \citealt{WoosleyWeaver1995}), while a stellar companion should display CNO cycle by-products.
Planet engulfment might increase stellar rotation \citep{Carlberg2010,Carlberg2012} that triggers internal 
lithium production \citep{SiessLivio1999} and/or could be connected with an increased mass loss that could 
be observable as an infrared excess \citep{delaRezaDrake1997} or as an additional component 
in the spectra of sodium lines (i.e. \citealt{Drake2002}). 

There is still less than a hundred Li-overabundant giants known. As only a small percent of giants reveal 
that characteristic, this type of star is usually discovered in large spectroscopic surveys. Studying Li abundance
for giants that are targets of a planet search gives unique opportunity to investigate Li abundance in connection with enhancement 
scenarios that involves a presence of stellar companions. 

In this paper, we report on a high-resolution spectroscopic search for Li enrichment in giant stars located along the RGB.
We use the extensive collection of high resolution spectra obtained for the Penn State-Toru\'n Centre 
for Astronomy Planet Search (PTPS) program \citep{Niedzielski2007, NW2008}. The full PTPS sample is composed of over 1000 evolved 
solar-type and intermediate-mass stars: giants, including the RGC stars \citep{Zielinski2012}, 
subgiants (Niedzielski et al. in prep), and evolved dwarfs (Deka et al. in prep). 
We have selected the most evolved subsample of PTPS stars, for which uniformly determined atmospheric 
and integrated stellar parameters are available from \cite{Zielinski2012}. Those are 348 stars (343 giants and five dwarfs), for which we 
determine lithium abundances. 
We aim to study various channels of Li-enhancement in giants and their connection 
to the stellar parameters in an attempt to better constrain the characteristics of the enrichment mechanism. 
  
The paper is organized as follows: in \S \ref{observations}, we describe the observational material that we collected for the analysis.
In \S \ref{Analysis}, we describe the methods that we applied to our sample 
for $\ALi$ and for the  determination of Al, O, Mg, Ti, and Ca abundances. We also present rotational velocities obtained 
with two independent methods and discuss uncertainties of results 
and abundances determined for Arcturus. 
In \S \ref{results-g}, we present results of $\ALi$ determinations and  investigate possible links between 
various stellar parameters and lithium abundance. In \S \ref{stars}, we discuss 
 individual objects with enhanced Li abundances in more detail, and in \S \ref{discussion}, we present discussion of the results.

\begin{figure}
   \centering
   \includegraphics[width=0.5\textwidth]{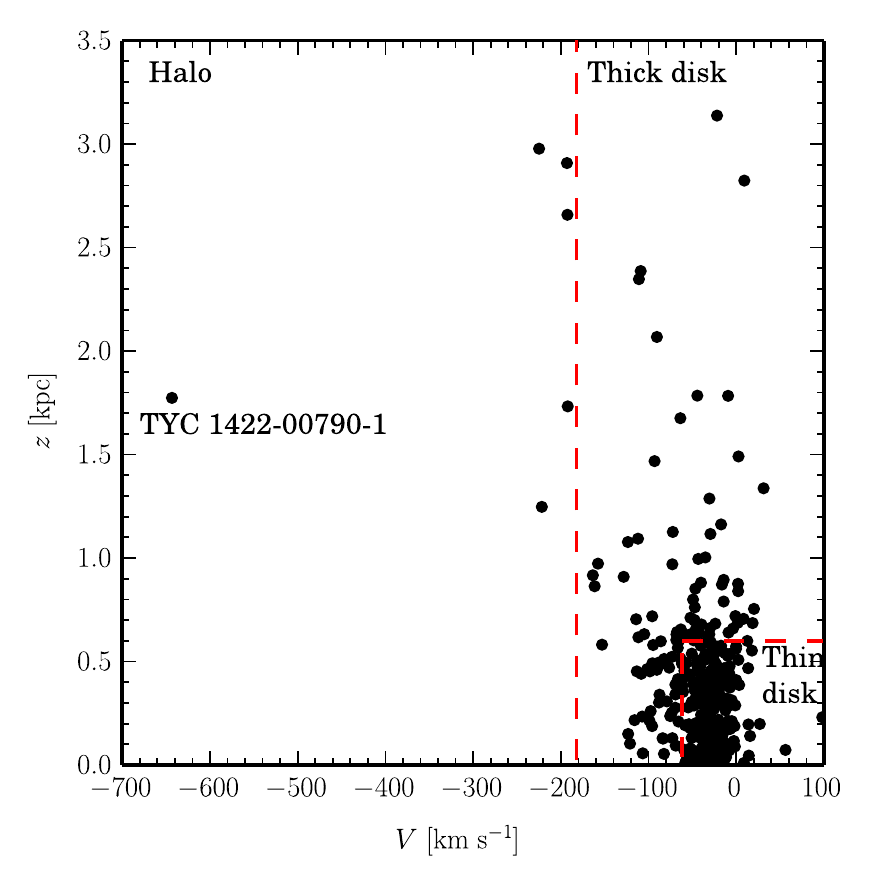}
      \caption{Galactic rotational velocities of stars with respect to the Galactic Center $V$ versus distances from the galactic plane for
       stars analyzed in this paper. We mark the location of the different galactic components, thin disk, thick disk, 
       and halo according to the \cite{Ibukiyama2002} criteria. The object TYC~1422-00790-1 is among the six  halo stars present in the sample.}
         \label{fig1}
         \end{figure}  

\section{Sample, observations and data reduction \label{observations}}

\subsection{Sample}

The sample presented in this paper contains 343 giants and five dwarfs out of the total $\sim1000$ stars sample
monitored in the PTPS project with the Hobby-Eberly Telescope (HET) for radial velocity variations using the high-precision
iodine-cell technique. The total PTPS sample is mainly composed of evolved low-mass and intermediate-mass stars:  
$\sim575$ giants, $\sim225$ subgiants, and  $\sim200$ slightly evolved dwarfs. We have selected
343 giant stars and five dwarfs for which \cite{Zielinski2012} has already derived effective temperatures, $\log g$, 
turbulence velocities and metallicities through a purely spectroscopic analysis for this paper. Stellar luminosities,  masses, 
and ages for those stars have been derived using fits to evolutionary tracks, and absolute radial 
velocities (RV) have been obtained with the cross-correlation technique (see \citealt{Zielinski2012}). 

A detailed  analysis revealed that the sample studied, presumably consisting  of RGC stars,
is actually composed mainly of regular giant stars evolving along the RGB ($62\pm3\%$).
Only $37\pm3\%$ of giants (126 stars) belong to the RGC, according to the extended criteria 
in luminosity, which is widened to account for uncertainties, as presented by \cite{Zielinski2012}; that is
i.e. 4700~K $\le \Teff \le 5100$~K 
and $1.3 \le log L/L_{\odot} \le 2$ (see \citealt{Jimenez1998, Taut09}). The sample also contains five dwarfs  ($1.4\pm1\%$).   

All stars have stellar parameters derived  in a homogeneous manner from the same set of  observations and instrument configuration. 
The stars have been shown not to be located in any particular region of the HR diagram which allows tests of the putative relation 
of Li overabundance and stellar evolution. Furthermore, all the stars included have reached the bottom of the RGB phase with the exception 
of two stars that are still in the subgiant phase and the five dwarfs mentioned previously. 

To check whether we have a mix of stellar populations in the sample, we have adopted the \cite{Ibukiyama2002} 
criteria and obtained the location of the stars in the galactic height versus galactic rotational velocity plane as shown in Figure \ref{fig1}. 
Distances from the galactic plane ($z_{max}$) have been determined using parallaxes from \cite{vanLeeuwen2007}, which are
available for 23\% of stars. For the remaining 77\% of objects, we have estimated distances using absolute magnitudes of \cite{Zielinski2012}. 
Assuming an average uncertainty in the determinations of $M_V$ given in that paper, the uncertainty for the estimated distances is $\sim100$~pc.
We computed space velocity components ($U, V, W$) following  \cite{Johnson1987} but used the 
J2000 epoch in the coordinates and the transformation matrix, 
stellar distances, and proper motions from the Tycho/Hipparcos Catalog \citep{vanLeeuwen2007}, and absolute RVs  from \cite{Zielinski2012} 
($\sigma RV_{CCF} = 0.041 \kms$). We then classified the stars into stellar populations using the
rotation velocity component in the Galaxy ($V$) and the distances from the galactic plane (see \citealt{Ibukiyama2002}).

We found (see Fig. \ref{fig1}) that most of our stars are disk giants (76\% of the sample), but we also found 
a non-negligible amount of stars that belong to the thick disk population, with 
high space velocity, and low [Fe/H] (22\%), according to the above criteria. We also found a small number of halo stars (six objects),
which represents  2\% of the sample.

\subsection{Observations and data reduction}
The observational material used in this paper are high resolution, high quality ($\mathrm{SNR}\ge200$ at 569~nm, typically) optical spectra
collected for the PTPS since 2004 with the HET \citep{Ram1998} located in McDonald Observatory. The telescope was equipped 
with the High Resolution Spectrograph (HRS) that is fed with a 2 arcsec fiber. The HRS was used in the R=60\,000 mode \citep{Tull1998}.
Observations were performed using the queue scheduled mode \citep{Shet2007}. The spectra consisted of 46 echelle orders 
recorded on the ``blue'' CCD chip (407.6-592~nm) and 24 orders on the ``red'' one (602-783.8~nm). 

Since the main goal of PTPS is to obtain high precision radial velocities, the observations were performed in two modes: 
(i) with the gas ($I_2$) cell inserted in the optical path and (ii) without the gas cell (GC1 and GC0, respectively). For every target, 
at least one high SNR ($\ge$300) GC0 spectrum  and a series of GC1 exposures were obtained.  The GC1 exposures 
contain the $I_2$ absorption spectrum imprinted in the first 17 orders of the ``blue'' spectra and are, therefore, of little use 
for further spectroscopic studies.  For the purpose of this paper, we used all available GC0 exposures and all  ``red'' CCD 
spectra where the $I_2$ lines are not present. Multiple ``red'' spectra per target allowed for good uncertainty estimates 
in the case of spectral lines redwards $\sim660$~nm. For spectral lines present 
in the ``blue'' CCD, typically only one measurement was possible.

The basic data reduction was performed using standard IRAF\footnote{IRAF is distributed by the National Optical Astronomy Observatories,
which are operated by the Association of Universities for Research in Astronomy, Inc., under cooperative agreement with
the National Science Foundation.} tasks and scripts developed for PTPS. The wavelength scale was corrected to the barycenter of the Solar System. 

Unfortunately, the HET/HRS flat-field spectra were sometimes contaminated with an emission spectral features near 
the $^{7}$Li $\lambda$670.8~nm and Na D1 and D2 lines. Those features may mimic the stellar absorption lines, 
depending on the actual RV of a star, and influence the abundance analysis,which leads to artificially increased values. 
As this features occurred in the flat-fields in a more or less random manner, all existing flat-field spectra had to be checked to avoid contamination.  

After rejecting all contaminated flat-field spectra, we used the uncontaminated ones from the same night if available.
Otherwise, we used data from the closest night. To avoid introducing additional noise in the reduced spectra in those cases, 
where only flat-field frames that are distant in time to the original data were available, we cross-correlated them first with 
each other to determine possible shifts on the CCD matrix.

Another problem we encountered in the ``red'' CCD reduction process was severe fringing for the data 
obtained before the new camera was installed in 2006. Fringe effects can be usually compensated by 
a proper flat-field correction. However, when the emission structure in the flat-field frames
 and the fringing appeared together, the replaced flat-field frames were not  sufficient to properly remove 
 the fringe pattern. In such cases, defective frames were rejected and excluded from further analysis.
 
To ensure that the instrumental effects described above were properly removed during data reduction,
spectra acquired with several other instruments were used to test the reliability of our results (\S \ref{Li-abundances}). 
To that end, seven high-resolution spectra (for five stars), which were obtained using the Spanish service time mode at the Instituto de Astrofisica de Canarias 
at La Palma observatory (Canary Islands, Spain), were obtained the night of the 24 of August 2012 with 
FIES \citep{FrandsenLindberg1999} at the Nordic Optical Telescope (2.56~m). The FIES spectra cover the wavelength range 
from 364 to 736~nm with a resolution of $\mathrm{R} \sim67\;000$. 
They were reduced using the advanced option of the automatic data reduction tool
FIEStool\footnote{See http://www.not.iac.es/instruments/fies/fiestool/FIEStool.html for details.}.  

We also used 15 spectra of another six targets from our ongoing program using 
the High Accuracy Radial velocity Planet Searcher-North (HARPS-N) spectrograph at 
the 3.58~m Telescopio Nazionale Galileo (TNG; \citealt{Cosentino2012}).
The HARPS-N, a near twin of the HARPS instrument in operation on the ESO 3.6 m telescope at La Silla (Chile), 
covers the wavelength range from 380~nm to 690~nm with a resolving power of $\mathrm{R} \sim115\;000$. The standard, automatic reduction pipeline was used.

One additional spectrum obtained at the MERCATOR telescope (1.2~m) with the HERMES spectrograph \citep{Raskin2011, Maldonado2013} 
was also used for our tests. The HERMES spectra have a resolution of  $\mathrm{R} \sim 85\;000$, cover the spectral range $\lambda380-900$~nm, 
and were automatically reduced using the data reduction pipeline of the instrument\footnote{See http://www.mercator.iac.es/instruments/hermes/ for details.}. 

All spectra used for independent verification of our HET/HRS data had S/N values between 50 and 150 around the lines of interest. 

We also compared the lithium abundance obtained for Arcturus based on the analysis of PTPS spectra 
with data available from the literature (see Section \ref{arctur}).

\section{Spectral analysis} \label{Analysis}

\begin{table*}
\centering
\caption{ $\ALi$ determinations based on spectra obtained with different instruments. All abundance values are given in dex.}
\begin{tabular}{llccccc}
\hline
Star &   & $\ALi_{HET}$ &$\sigma_{s,HET}$& $\ALi$& $\sigma_s$ &Instrument\\
\hline\hline
TYC 0870-00114-1 & HIP 57428  & $<0.63$ &      & 0.89 &      & HERMES \\ \hline
TYC 0435-03332-1 & BD+02 3497 & 1.44    & 0.01 & 1.42 & 0.01 & FIES \\
TYC 1422-00790-1 & BD+20 2457 & 0.32    & 0.01 & 0.29 & 0.02 & FIES \\
TYC 3226-02285-1 & HIP 111944 & 0.73    & 0.01 & 0.74 & -    & FIES \\
TYC 3304-00101-1 & HD 17092   & $<-0.10$&      &$<-0.26$ &   & FIES \\
TYC 4006-00980-1 & HD 240210  & $<-0.17$&      &$<-0.80$ &   & FIES \\
\hline
TYC 0684-00553-1 &            & 3.06 & 0.02 & 3.16 & 0.01 & HARPS-N  \\
TYC 3226-01219-1 &HD 216016   & 1.10 & 0.01 & 1.15 & 0.01 & HARPS-N \\
TYC 3304-00090-1 &  BD+48 740 & 1.90 & 0.01 & 1.94 & 0.01 & HARPS-N \\ 
TYC 3917-01107-1 & HD 238914  & 1.83 & 0.01 & 1.85 & -    & HARPS-N \\
TYC 3993-01850-1 &            & 1.75 & 0.07 & 1.80 & 0.02 & HARPS-N\\
TYC 4421-01779-1 & HD 160723  & 1.04 & 0.03 & 1.07 & -    &HARPS-N \\
\hline 
\end{tabular}
\label{tab1}
\end{table*}

\subsection{Abundances determination} 
 
\begin{figure}
   \centering
   \includegraphics[width=0.5\textwidth]{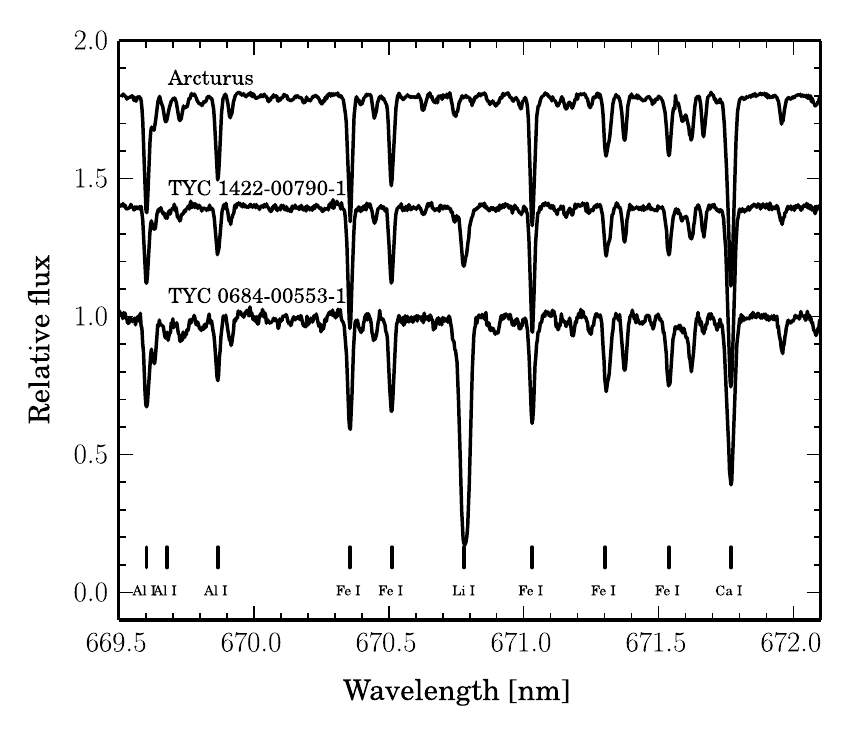}
   \caption{Spectral region used for the Li abundance determination using the SME package.
   The HET spectra of three representative stars are shown.}
   \label{fig2}
\end{figure}

The spectral analysis was performed with the Spectroscopy Made Easy package (SME, \citealt{SME1996}). 
This tool allows us to fit a synthetic spectrum for a star of given spectral type and RV to the observed one. 
As output, one obtains abundances of the chosen elements and refined stellar parameters. 
The SME the assumes local thermodynamical equilibrium and plane parallel geometry but ignores 
the effects of stellar magnetic fields and mass loss. 
For constructing synthetic spectra, we used 1D-LTE plane-parallel atmosphere models from ATLAS9 by \cite{Kurucz1993}. 

The SME requires a list of atomic data for lines to be  analyzed  in the format provided by
the Vienna Atomic Line Database (VALD, \citealt{VALD1999}) as an input parameter. For every spectral range to be studied, we 
extracted a starting  list of lines representative of an ''average'' giant star in our sample from VALD, 
which is a star with $\Teff=4700$~K, $\log g=2.8$, and $\vmic=1.4$. 
That list was later revised and updated, as described in sections \ref{Li-abundances}, \ref{oxygen}, and \ref{alpha}.

Stellar parameters that are required as the input of  SME  ($\Teff$, $\log g$, $\vmic$, [Fe/H], and RV) were 
adopted directly from \cite{Zielinski2012}  and were kept as fixed in analysis.
The [Fe/H] was assumed to be  the closest proxy for the stellar metallicity [M/H].
When solving for individual abundances, the SME uses [M/H] as a scaling factor  to interpolate between models of atmospheres.
 
There are three major factors that influence spectral line broadening without changing its depth: instrumental profile, 
macroturbulence velocity $\zeta$, and rotation velocity. We defined the instrumental profile as a Gaussian function parametrized 
by spectral resolution.  In case of PTPS spectra, this is 60~000. It is assumed, that broadening profiles for macroturbulence 
and rotation velocity are Gaussian in shape in SME, but the input value for $\zeta$ is actually the radial tangential macroturbulence parameter, 
rescaled to the Gaussian macro turbulence by using a factor of $\sqrt{2}$ .
For giants, rotation velocity contributes less to line broadening than macroturbulence. Hence, we adopted $v \sin i = 0$ 
and solved only for macro turbulence.We present our rotation velocity determination method in Section \ref{RotVel}.

\subsection{Lithium abundance}\label{Li-abundances}

The Li  abundance analysis was performed using only the order of the ``red'' HET/HRS spectra, which contains the spectral 
range from $\lambda$670.3 to 671.1~nm (see Fig.  \ref{fig2}). There are no $I_2$ lines present in that  range, 
and  we used all available GC1 spectra. Their actual number depends on the star status in the PTPS observing program
and can range from 2 to about 100 per star. We analyzed 4750 spectra in total with 14 spectra  per star on average. 

With SME, we fitted  the $^7$Li line at $\lambda$670.8~nm with several lines of C, N, V, Sm, and Fe that are 
present in that spectral region and the macroturbulence velocity. All initial abundance values are solar, 
except for C and N, which abundances we expect to be processed by first dredge up. We therefore  adopted 
abundances typical for a red giant:  C/H = -4.01 and N/H=-4.07 \citep{Iben2012}. 

We replaced the initial VALD list of lines with that  from \cite{Carlberg2012} but we expanded it with two Fe lines 
at $\lambda$670.36 and 671.03~nm, for which we used line data from VALD ( $\log gf = -3.16, \; -4.88$, respectively). 
For vast majority of giants analyzed here, Li lines, as well as nearby CN and elemental lines
are weak, blended structures,  therefore the two Fe lines were  included to control RV shift 
(we allowed SME for small adjustments of RV) and line broadening. 

To check if  the instrumental effects discussed in Section \ref{observations} were properly removed,
we analyzed spectra of 12 stars from our sample obtained with other telescopes and instrumental configurations. 
Our Li abundances obtained from HET spectra and those obtained from HERMES, FIES, and HARPS-N 
spectra are presented in Table \ref{tab1}, where column (1) lists the target TYCHO name and column (2) an alternate identification; 
columns (3) and (4) provide  lithium abundances derived using HET spectra  and their  uncertainties 
($\sigma_s$, which describe spectra quality only, whenever multiple spectra per target were available, see also Section \ref{errors}). 
Columns (5) and (6) list the lithium abundances 
and uncertainties obtained from instruments listed in column (7). 
Of all stars presented in Table \ref{tab1}, only 
one,  TYC 3226-02285-1, was observed during  nights 
with all flat-field frames not containing emission lines. For the rest of stars, the number of observing nights for which 
all flats were contaminated by the emission lines varied from 1 for TYC 3226-01219-1 to 18 for TYC 3917-01107-1.
The Li abundances determined from spectra obtained with different instruments are very similar, 
which are typically within estimated uncertainties of individual determination, to those obtained from HET/HRS spectra. These comparisons give us confidence on the validity 
of the procedure for the proper removal of instrumental effects in the HET data and, therefore, 
on the reliability of the abundances derived from HET/HRS data. 

Another important issue to consider is that the lithium line at $\lambda$670.8~nm, which is 
the base of our analysis, may be subject to several non-LTE processes, such as 
photon loss, ultraviolet overionization, and bound-bound pumping (\citealt{Carlsson1994} and references therein). 
All these processes are ignored in SME analysis and thus need to be taken into account separately. 
We applied the non-LTE corrections  provided by \cite{Lind2009} to our LTE  lithium abundances 
for a star of  given $\Teff,\; \log g,\; \vmic$ and [Fe/H]. 
For three stars, $\vmic$ fell outside the range of \cite{Lind2009} grids, and we applied a non-LTE correction 
for the maximum available value of $2\kms$. For 63 objects, our LTE Li abundances 
were below the minimum value, for which curves of growth are defined in \cite{Lind2009}. For those stars, 
we adopted non-LTE correction for the lowest LTE value provided.
For most of the stars in our sample, the non-LTE corrections are on the order of 0.2 dex. 

\subsection{Oxygen abundance}\label{oxygen}

Oxygen abundances, [O/H] were obtained from the analysis of  the $\lambda$777.2, 777.4, and 777.5~nm triplet 
and, separately, from three lines at $\lambda$557.7, 630.0, and 636.3~nm.

 Atomic data  for each line of the triplet were adopted directly from VALD ($\log gf = 0.369, 0.223, 0.001$, respectively). 
Those spectral features appear in two of the HET/HRS ,,red'' spectra echelle orders 
with slightly different signal-to-noise ratio. We separately extracted the spectral 
lines from both orders, doubling the number of input files introduced into 
SME for the analysis. The spectral range of the oxygen triplet is beyond any
 possible contamination from the $I_2$ absorption, so we used all  available GC1 
 spectra for every star (between two and 100 spectra, 14 on average). 
 The analysis was performed separately for each line of the triplet. Since the oxygen 
 triplet is strongly affected by non-LTE effects, we applied the non-LTE corrections 
 by \cite{Ramirez2007} to the abundances derived using SME.  
 For 59 stars with $\Teff<4400\mathrm{K}, \log g<2$ and/or $[\mathrm{Fe/H}]>0.4$ 
 stellar parameters fell out of the non-LTE grids and extrapolated non-LTE values were used.

For the three oxygen lines at $\lambda$557.7, 630.0, and 636.3~nm,
$\log gf$ data were updated using line list by \cite{RamirezAllendePrieto2011}.
Those lines are less affected by non-LTE effects but blended 
($\lambda557.7$nm with Y, $\lambda630.0$ with Ni) and available only
in limited number of GC0 spectra.  No non-LTE correction was applied to oxygen
abundances obtained from those three lines.

Both oxygen abundances agree within uncertainties 
(see Section \ref{errors} for more detailed discussion on uncertainties estimates).
Hence, the final oxygen abundance presented is the average of those two values.

\subsection{Aluminium, magnesium, titanium, and calcium abundances}\label{alpha}

Abundances for all four elements were determined using narrow $0.1-0.22$~nm ranges 
of spectra, centered on unblended lines. We used Al, Mg, Ti, and Ca lines in the range  $\lambda$520--660~nm 
selected by \cite{RamirezAllendePrieto2011}. We also updated the initial VALD line list using $\log gf$ from that paper. 
Lines of Mg, Ca and Ti are present  in spectral regions strongly affected  by $I_2$ lines. Hence, we used only
GC0 spectra  for SME analysis.  Lines of Al are not affected by iodine spectra. Therefore, both GC0 and GC1 
spectra were in use.

We did not apply non-LTE corrections to the obtained abundances.  
Non-LTE corrections of Al,  Mg, Ti and Ca are usually calculated for metal-poor stars
 (halo stars and stars in globular clusters) or are for far infrared lines (beyond $\lambda$1000~nm). Most of our stars 
 fall out of the range of stellar parameters provided in the available grids (see e.g. \citealt{Spite2012} for Ca). 
 Non-LTE corrections for solar-metallicity giants are sorely needed \citep{LuckHeiter2007}.
 
\subsection{Abundances uncertainties}\label{errors}

There are several sources of uncertainties in abundance analysis.
One of them are uncertainties connected with the quality of spectra (variable signal-to-noise ratio, 
continuum fitting, and changes in the instrumental profile). 
The availability of multiple spectra for every single star represents an advantage 
for the uncertainty calculation.
Next are uncertainties of line-to-line variations whenever analysis includes more than one line per element.
Uncertainties introduced by  stellar parameters adopted in analysis should also be taken into account.
In the case of Li and O, application of non-LTE corrections also contributes to the final uncertainties.

The contribution to the uncertainties of spectra quality and line variations, $\sigma_s$ were estimated from the standard deviation 
obtained from multi-epoch observations  and scatter in abundances due to multiple lines analysis for a particular element. 
The average uncertainties for stellar parameters adopted from \cite{Zielinski2012} are: $\Delta \Teff=40 \mathrm{K}, \; \Delta \log g =0.15,\;
\Delta[\mathrm{Fe/H}]=0.07$, and $\Delta \vmic =0.24$. We estimated their contributions to the final uncertainties, 
($\sigma_T, \sigma_g, \sigma_{Fe} and \sigma_{\xi}$, respectively) by varying the input parameters  by those values. 
The results are presented in Table \ref{tab2}.
The total uncertainty is a quadratic sum of all sources of uncertainties listed. As the number of individual determinations 
was typically small ($<30$), the standard deviation is multiplied by the t-Student\footnote{Small-number t-Student statistics 
by William Sealy Gosset (,,Student'') was first described  in  \cite{Student1908}} coefficient for confidence level of 68$\%$ (i.e. 1 $\sigma$) - $\alpha$=0.32 

For Li abundances, the uncertainty was calculated from multiple LTE determinations.
Uncertainty estimates for the non-LTE corrections for Li were not available.
The final Li uncertainties are therefore those resulting from LTE  analysis.

For oxygen triplets, multiple abundance determinations were available.
The final uncertainty was then calculated by averaging the uncertainties obtained for each single line of the triplet.
The procedure to calculate the non-LTE correction for O did not take into account the uncertainty in the abundance determination 
as an input parameter, but it returned an line-to-line uncertainty.
Thus, for oxygen triplet abundances, the $\sigma_s$ is a quadratic
sum of the uncertainties resulting from the instrumental effects and the uncertainties from non-LTE corrections.

For the three oxygen lines at $\lambda$557.7, 630.0 and 636.3~nm, the abundance 
uncertainty was calculated from the scatter in three determinations. 

One should mind the differences for oxygen: the triplet lines 
are more sensitive to stellar parameters, 
while the line-to-line uncertainties for the other three LTE lines and poor statistic 
(high t-Student coefficients) strongly contributed to 
total uncertainty. In both cases, the average uncertainty for oxygen abundance is 0.2~dex.

\begin{table}
\centering
\caption{Uncertainties in abundance analysis propagated from stellar parameters errors.}
\begin{tabular}{c|cccc}
\hline
&$\sigma_T $&$\sigma_g$ &$\sigma _{Fe} $&$\sigma_{\xi}$ \\
\hline\hline
Li & 0.06 & 0.14 & 0.13 &0.03 \\
 O$_t$ & 0.08 &0.09& 0.10 &0.10\\
 O & 0.01& 0.04&0.02 &0.01 \\
Mg &0.05 & 0.02 &  0.08 & 0.06\\
  Al &  0.03 & 0.01 &  0.07 & 0.04\\
 Ca &  0.04 & 0.07 &  0.06 & 0.10\\
   Ti &  0.07 & 0.02 &  0.03 & 0.10\\
\hline
\end{tabular}
\label{tab2}
\end{table}

%%%%%%%%%%%%%%%%%
\subsection{Rotational velocities}\label{RotVel}

\begin{figure}
   \centering
   \includegraphics[width=0.5\textwidth]{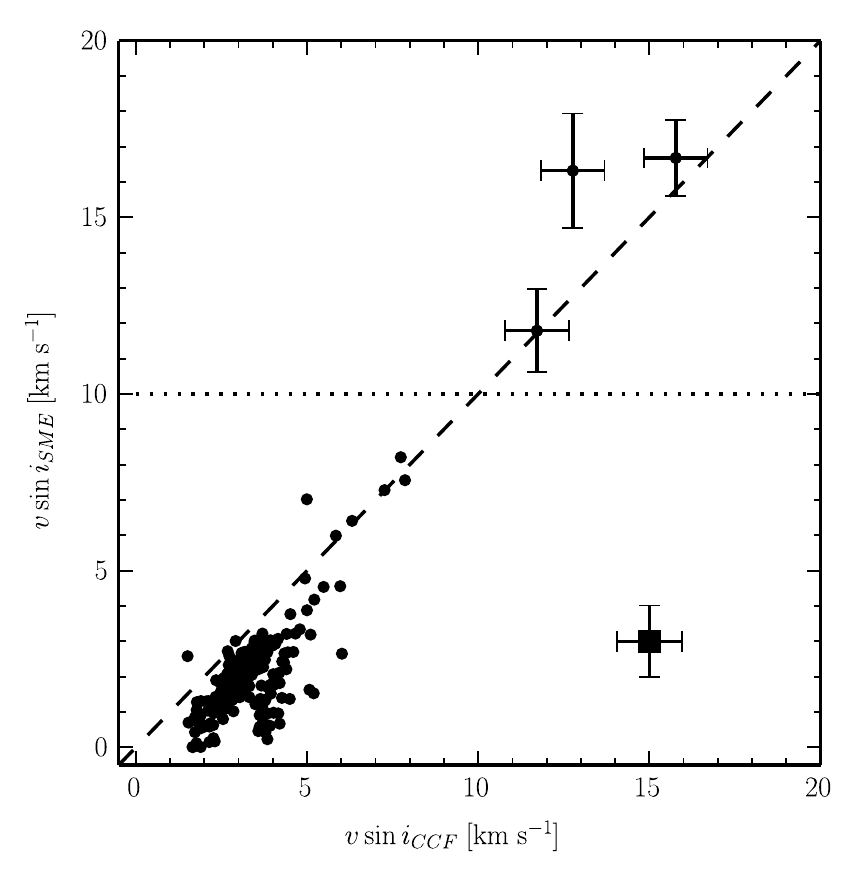}
   \caption{Stellar rotation velocities obtained using the cross-correlation technique - $v \sin i_{\mathrm{CCF}}$ 
   and those obtained from SME synthetic spectrum fitting -  $v \sin i_{\mathrm{SME}}$.
    The dashed line marks a one-to-one relation. The dotted one separates the fast and slow rotators. 
    In  the lower right corner, an average uncertainty in $v \sin i$ determination  (for slow rotating stars) for both methods is presented.}
   \label{fig3}
\end{figure}

In the case of giants, rotational velocities are expected to be smaller than macro turbulence velocities and, 
hence, contribute less to the total line broadening. We have, however, estimated them within  
our spectral analysis, as rotation could be related to the lithium overabundances; the nature of which we are trying to understand.
We obtained the rotational velocities using two independent methods: by measuring the line width through
cross correlation as described in \cite{Nowak2013} and by SME spectral modeling. 
In both methods, the macroturbulence velocity $\zetahm$
was adopted from the \cite{HekkerMelendez2007} calibrations for every star, which is an effective temperature
 and surface gravity combination. We assumed, that rotational velocity can be expressed as:
\begin{equation}\label{vrot}
v\sin i = \sqrt{\beta^2 - C\zetahm^2},
\end{equation}
 where $\beta$ stands for the total line broadening (assumed to be Gaussian) and $C$ is a factor
scaling the radial-tangential $\zetahm$ values based on work by Gray and collaborators with Doppler 
 shift distributions for macroturbulence velocity \citep{Massarotti2008, Fekel1997}.
 If the respective  $\zetahm$ from \cite{HekkerMelendez2007} 
  was higher than the  total  line broadening, we followed the widely accepted procedure and adopted $\zetahm$ for lower luminosity class.

The cross-correlation functions (CCFs) were computed by cross-correlating the stellar spectra (GC0) with a 
numerical mask consisting of 1 and 0 value points with the non-zero points corresponding to the positions of the
 stellar absorption lines at zero velocity. The mask was constructed from a synthetic K2 star spectrum 
 of \cite{Kurucz1993} ATLAS9 and contained about 300 lines. The CCFs were computed in all the 17 HRS orders, 
 which cover $\lambda$504-592~nm spectral range and are used for RV measurements over a wide $\pm 30\kms$ range
after correcting the spectra for their absolute radial velocities and shifting them to the Solar System barycenter \citep{Stumpff1980}. 
Finally, the 17 CCFs were added together in the wavelength scale to form the integrated CCF.
To determine $v\sin i$, we followed the procedure from  \cite{Carlberg2011} but assumed $C=0.5$ and subtracted the instrumental profile of $5 \kms$.

In SME analysis, we measured line broadening by assuming that it stems from macro turbulence only. This allowed us to
obtain a parameter which should be interpreted as an upper limit for the radial-tangential macro turbulence $\beta_{\zeta}$. 
Assuming a theoretical macroturbulence velocity $\zetahm$
 from  the \cite{HekkerMelendez2007} calibrations for every star,  
we modyfied eq. (\ref{vrot}) and estimated the rotation velocity $v\sin i$ as:

\begin{equation}\label{vrot2}
v\sin i = \sqrt{C(\beta_{\zeta}^2 - \zetahm^2)}.
\end{equation}

Uncertainties in the rotational velocities were calculated by
taking dispersions in the total broadening and $\zetahm$ fitting into account.

\begin{table*}
\centering
\caption{Comparison of $\alpha$-elements abundances for Arcturus.}
\begin{tabular}{c|ccc|c|c}
\hline

&\multicolumn{3}{c|}{This work} &   \cite{RamirezAllendePrieto2011} &\cite{Smith2013} \\
x&A(x)$_{\odot}$ &A(x) &  $[$x/H$]$ & $[$ x/H$]$ &A(x)\\
\hline\hline
 O & 8.62 & $8.29 \pm 0.14$ & $-0.28 \pm 0.14$ & $-0.02 \pm 0.03$ & $8.64 \pm 0.04$\\
Mg & 7.49 & $7.14 \pm 0.14$ & $-0.35 \pm 0.14$ & $-0.15 \pm 0.03$ & $7.15 \pm 0.08$\\
Al & 6.33 & $6.14 \pm 0.10$ & $-0.19 \pm 0.10$ & $-0.18 \pm 0.03$ & $6.16 \pm 0.01$\\
Ca & 6.27 & $5.73 \pm 0.14$ & $-0.54 \pm 0.14$ & $-0.41 \pm 0.04$ & $5.84 \pm 0.07$\\
Ti & 4.86 & $4.55 \pm 0.24$ & $-0.32 \pm 0.14$ & $-0.25 \pm 0.05$ & $4.59 \pm 0.07$\\
Fe & 7.41 & 6.89$^a$ & -0.52$^a$ & $-0.52 \pm 0.04$ & $6.98 \pm0.04 $\\

\hline
\end{tabular}
\tablefoot{$^a$  abundance value adopted from \cite{RamirezAllendePrieto2011}.}
\label{tab3}
\end{table*}

Figure \ref{fig3} shows that the rotational velocities obtained with the two methods are generally in good agreement.
Three rapid rotators, which are defined usually as stars with $v\sin i >10\kms$ \citep{Fekel1997} are confirmed. These are:
TYC 3930-01790-1, as identified already by \cite{Massarotti2008}, 
and  TYC 3676-02387-1 and TYC 3318-00020-1, which are both detected by \cite{Nowak2012}. 

For stars with $v \sin i<5 \kms$, the SME analysis revealed rotational velocities systematically lower by $\sim1\kms$ as
compared to those obtained with CCFs, which is probably due to different methodology applied to their determination.
A small group of stars, slow rotators, for which $v \sin i_{\mathrm{CCF}}-v \sin i_{\mathrm{SME}}\gtrsim2\kms$ is also present in Fig. \ref{fig3}.
These are objects, for which  $\beta<\zetahm$ and realistic $v\sin i_{\mathrm{CCF}}$ determination  required a change in luminosity class
to obtain  $\zetahm$ from \cite{HekkerMelendez2007} calibrations,
while the correct and hence higher $\zetahm$ could be applied to calculate $v \sin i_{\mathrm{SME}}$.

\subsection{Consistency check}

\subsubsection{Arcturus abundances}\label{arctur}

To test the aforementioned methodology, we applied it to  Arcturus. Three very good quality ($\mathrm{SNR}\sim300$) HET/HRS spectra in
 standard PTPS configuration were obtained on 16 Jan, 27 Jan and 20 Feb 2012. 
 Our Li abundance, $\ALi_{LTE}<-0.74$, agrees very well 
 with the determinations of  \cite{Carlberg2012} with $\ALi<-0.73$, \cite{Reddy2005} with $\ALi<-0.6$ 
 and \cite{Brown1989} with $\ALi<-0.8$, despite that all flats were contaminated 
 with emission lines.
 In  Table \ref{tab3} we present  $\alpha$-elements abundances for Arcturus obtained here  and compare 
them to recent results of several other authors. We also present the adopted solar values used in calculation of [X/H] through this paper in Table \ref{tab3}.
For Al,  Ca, and Ti, our abundances are in very good agreement with those obtained by \cite{RamirezAllendePrieto2011} and \cite{Smith2013}. 
In case of Mg, we practically obtained identical value as \cite{Smith2013}. 
Our oxygen abundance is lower by 0.3~dex  than those presented by \cite{RamirezAllendePrieto2011} and \cite{Smith2013}.
We note, however, that the oxygen abundance for three LTE lines is $[\mathrm{O/H}]=-0.24 \pm 0.09$  and the oxygen 
abundance discrepancy is not a result of extrapolated non-NLTE correction for the triplet lines.

Arcturus is a slow-rotating star. Our rotational velocity of $v \sin i_{\mathrm{SME}} =2.7 \pm0.6 \kms$
agrees very well with those of \cite{Carlberg2012} at $2\pm 1.5\kms$ and  \cite{Carney2008} at $2.4\pm 1 \kms$.

\subsubsection{$\alpha$-elements abundances vs. metallicity}

\begin{figure}
   \centering
   \includegraphics[width=0.45\textwidth]{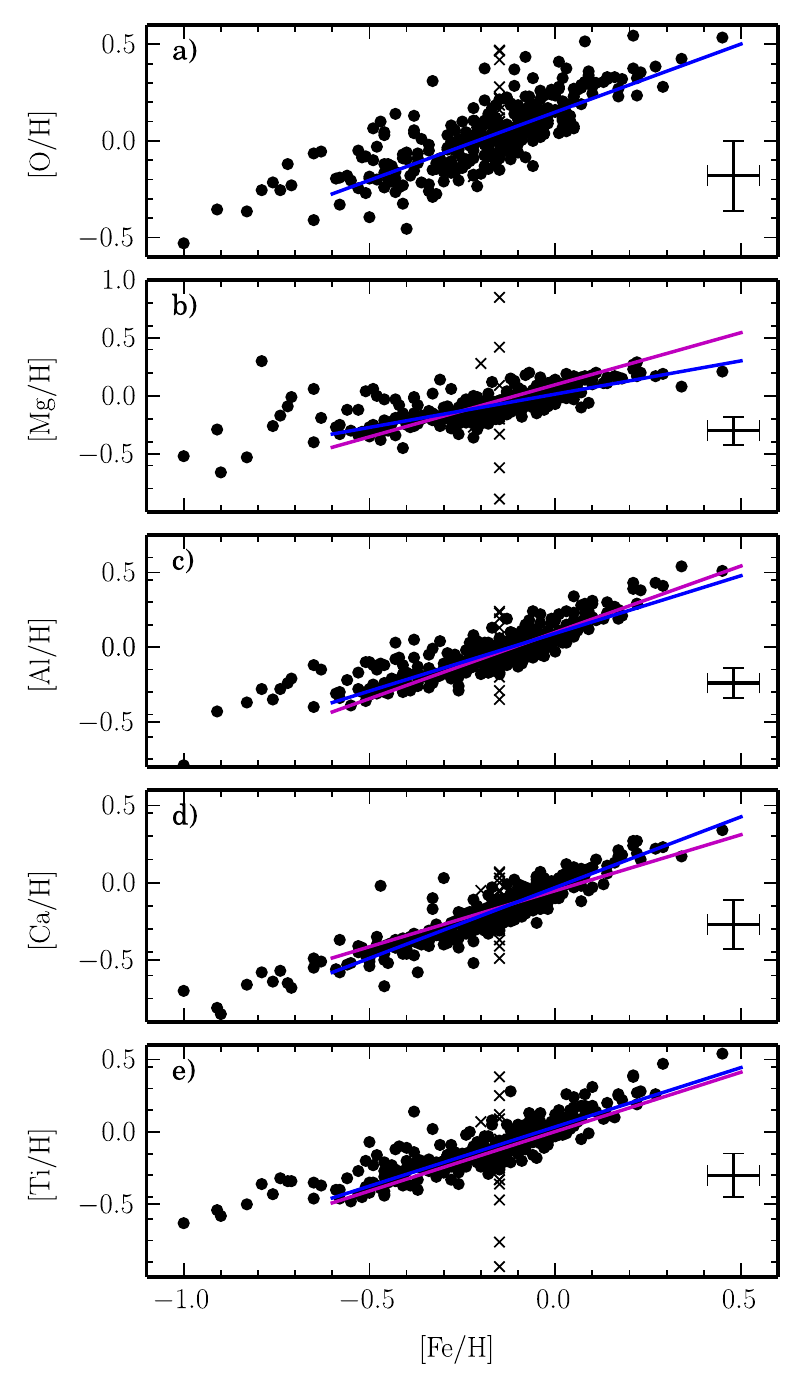}
      \caption{Abundances for the complete sample of giants (crosses denote  stars with uncertain atmospheric parameters). 
      The error bars in the lower right corners are the mean uncertainties for [Fe/H], as adopted from \cite{Zielinski2012} and
      for abundances of a given element obtained in this work. Blue lines represent linear
       regression fits with coefficients presented in Table \ref{tab4}, while magenta lines are
        fits of \cite{LuckHeiter2007}.  }
         \label{fig4}
         \end{figure}

Abundances of O, Mg, Al, Ca,  and Ti as a function of [Fe/H] are shown in Figure \ref{fig4}, where we also present 
 trends of [X/H] with [Fe/H] as obtained by \cite{LuckHeiter2007} and fits to these relations.
Many of stars with [Fe/H]$<$-0.6 are a thick disk or halo population objects that seem to form a different 
relation. Therefore, we fitted only stars with  [Fe/H]$>$-0.6 to compare our results with  \cite{LuckHeiter2007}. 
Coefficients for linear fits to our data for each element are presented in Table \ref{tab4}. General relations between elemental abundances  
and [Fe/H] agree  with those of \cite{LuckHeiter2007}, and we are confident that our $\alpha$-elements abundances are correct. 

\begin{table}{}
\centering
\caption{Coefficients for linear fits of [Fe/H] vs. [X/H] trend. ($\mathrm{[X/H]} = \mathrm{slope} \times \mathrm{[Fe/H]}+ \mathrm{intercept}$).}
\begin{tabular}{c|cc}
\hline
X&slope& intercept\\
\hline\hline

 O  & $0.705 \pm 0.103$ & $ 0.149 \pm 0.024$\\
Mg & $0.574 \pm 0.029$ & $ 0.015 \pm 0.007$\\
Al   & $0.771 \pm 0.064$ & $ 0.092 \pm 0.015$\\
Ca & $0.915 \pm 0.030$ & $-0.031 \pm 0.007$\\
Ti   & $0.818 \pm 0.036$ & $ 0.035 \pm 0.009$\\
\hline
\end{tabular}
\label{tab4}
\end{table}

%%%%%%%%%%%%%%%%%%%%%%%%%%%%%%%%%%%%%%%%%%%%%%%%%%%%%%%%%%%%%%%%%%
\section{Results - general}\label{results-g}
      \begin{figure}
   \centering
   \includegraphics[width=0.5\textwidth]{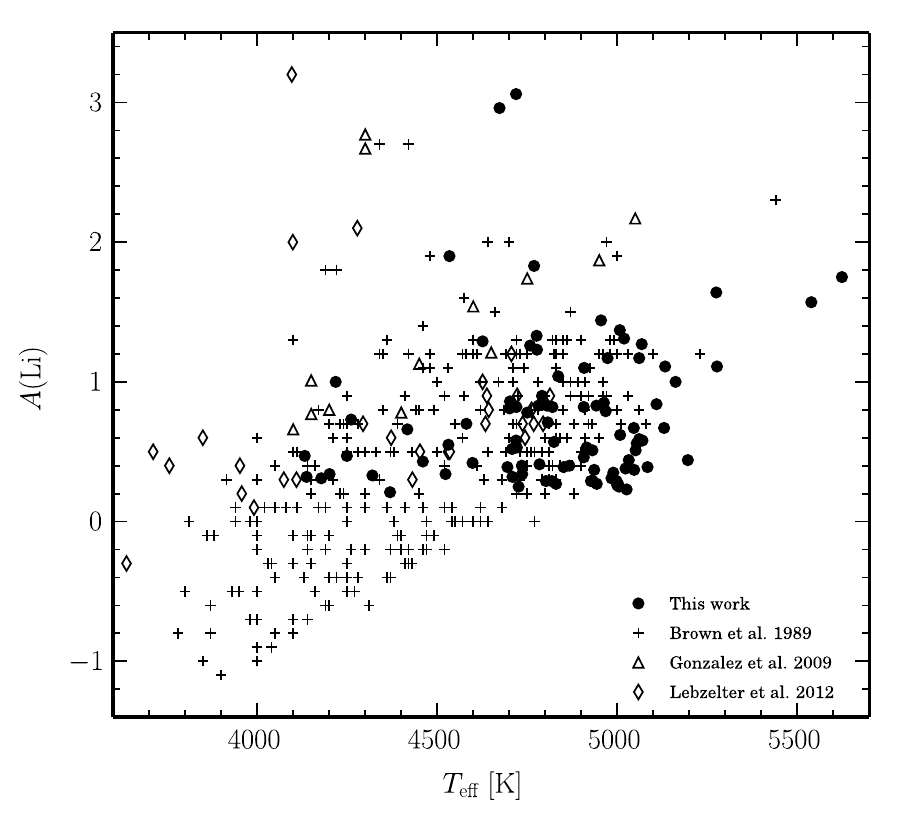}
      \caption{Comparison between the lithium abundances and stellar effective temperatures obtained in this work with 
      those obtained by \cite{Lebzelter2012}, \cite{Gonzalez2009}, and \cite{Brown1989}. For consistency, our $\ALi_{LTE}$ determinations  
      are presented for 92 stars (5 dwarfs, 2 subgiants  and 85 giants) with detected Li.}
         \label{fig5}
    \end{figure}

Our abundance determinations for Li, O, Mg, Al, Ca, and Ti 
as well as rotational velocities  with their derived uncertainties for all investigated objects
are presented in Table \ref{tab5}.
For each of the 348 stars analyzed, column (1) gives the target name according to the TYCHO catalog; 
columns (2-5) list the average rotation velocities obtained with the two methods and their uncertainties.
In column (6), we give the lithium abundance (or its upper limit) based on multi-epoch observations; 
column (7) contains the lithium abundance after non-LTE correction.
Column (8) provides the uncertainty obtained for $\ALi$.
Columns (9) to (18) list the O, Mg, Al, Ca, Ti abundances and their estimated uncertainties, 
and column (19) contains  information on the stellar subsystem to which each given star 
belongs according to \cite{Ibukiyama2002} criteria. The last column (20) provides additional information about particular object.
Although the table includes the results for five dwarf stars identified by \citealt{Zielinski2012}, 
they have been excluded from further analysis. 

Regarding the Li abundance, which is the main focus of this paper, we identified 92 stars with 
detectable Li lines in our sample. Most of $\ALi$ abundances are based on multi--epoch sets of spectra,
and each determination was found to be consistent within an estimated $\sigma\ALi$ level. 
Among the 92 detections, we identified 12 giants with large $\ALi$ abundances and two with lithium 
abundances close to the meteoritic value. For 271 stars with $\ALi<0$ or $\ALi\lesssim \sigma{\ALi}$, 
we provide only upper limits defined as the maximum value of $\ALi$ for a set of spectra for a given star. 

A comparison between the lithium abundances obtained in this paper and those available from the 
literature is presented in Fig. \ref{fig5}. The results of extended spectral 
analysis of \cite{Brown1989} based on high resolution spectra for 644 field  stars, 
and data from more recent studies by \cite{Gonzalez2009} and \cite{Lebzelter2012}, which are
focused on stars in the Galactic Bulge, are plotted together with our results. All available Li abundances, 
including those presented here, show an identical tendency regarding lithium
 abundances and $\Teff$; cooler stars reach lower Li abundances. Our sample
  is comparable to that analyzed by \cite{Brown1989}: it covers similar range in $\log g$ 
  and $\Teff$ and focuses on field stars. The main difference is the more restricted 
  selection of stars in the PTPS. The sample presented here  does not include stars 
  with $\Teff\lesssim4000$ nor are masses greater than $\sim3.5 \Msun$, while the 
  \cite{Brown1989} sample does. These limitations are the consequence of the
  observational strategy which is required for searching planets 
  around evolved stars. 
  
\begin{figure}
   \centering
   \includegraphics[width=0.5\textwidth]{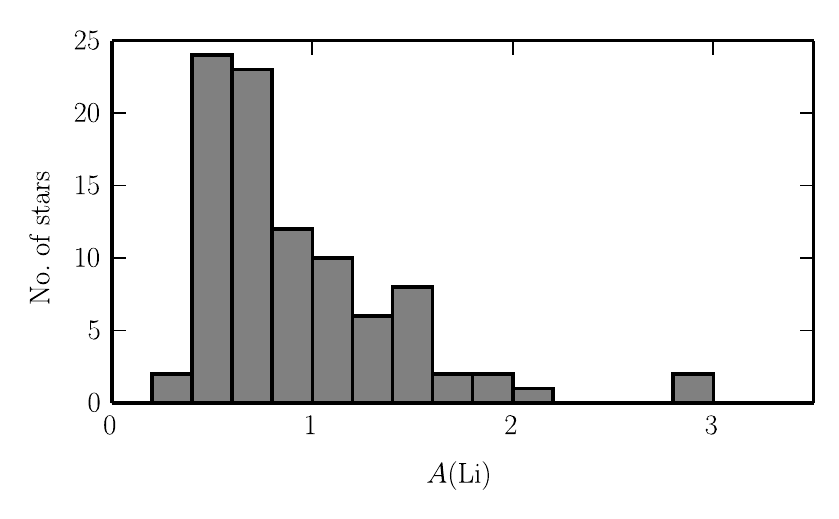}
   \caption{Histogram of $\ALi$ for 92 stars with Li detections. We note a jump in the number of stars  at $\ALi>1.4$. }
   \label{fig6}
\end{figure}

  In Fig. \ref{fig6}, we present a distribution of obtained $\ALi$. In addition to a gradual decay, 
  we observe a sudden increase of stars with $\ALi>1.4$ and a few super-Li giants.
  
\subsection{Giants with lithium on the Hertzsprung-Russel  diagram}

\begin{figure}
   \centering
   \includegraphics[width=0.5\textwidth]{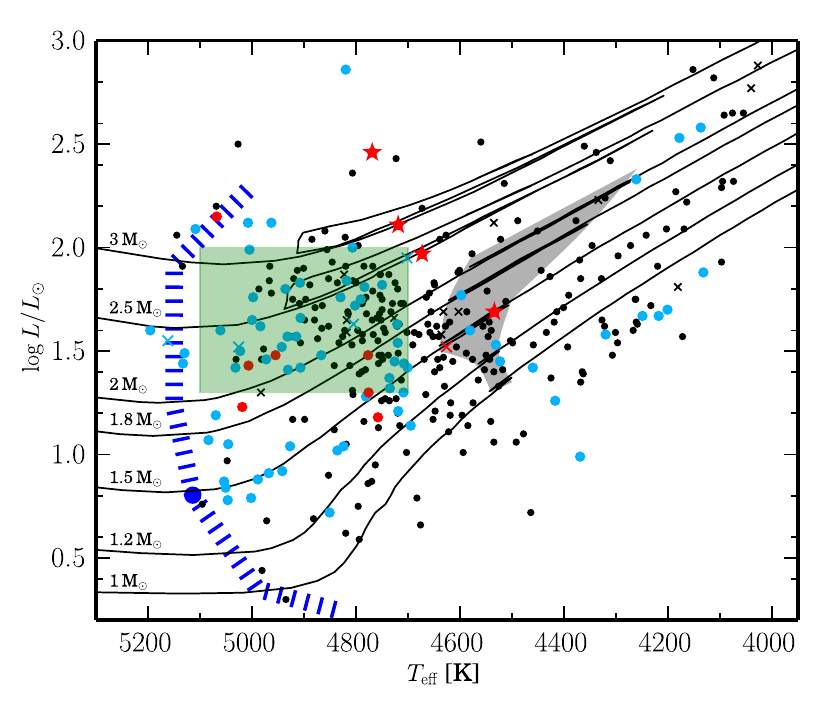}
   \caption{HR diagram showing the location of all the PTPS stars studied in this paper. Colors of points denote Li abundance. Red are the most Li-rich objects 
   (the red stars denote the four most Li-rich giants). Stars denoted in blue have moderate Li level, and black is for Li-poor stars. 
   Crosses denote  stars with uncertain atmospheric parameters. 
   The blue stripes denote the beginning of RGB. 
   The gray area is LFB region, and green box denotes RGC. 
   Evolutionary tracks from \cite{Bertelli2008} for $1-3\Msun$ stars with solar metallicity are plotted as solid lines.}
   \label{fig7}
\end{figure}

The location of our stars in the Hertzsprung-Rusell (HR) diagram  is shown in Fig.~\ref{fig7}, 
where evolutionary tracks of  1 to $3\Msun$ stars   and solar metallicity  from \cite{Bertelli2008} are  presented as well.
 
To investigate the Li abundance of our sample of PTPS giants, we have divided 
the total sample (323 stars) in three groups:  (1) stars with only upper limits or no lithium detections 
(241 objects, group A, black symbols in Fig.~\ref{fig7}); (2) stars with detected Li and  $\ALi_{NLTE} <1.4$ 
(71 objects, group B, blue symbols); and (3) giant stars with Li--abundances 
 $\ALi_{NLTE} > 1.4 $ (11 objects, group C, red symbols).

In addition to the five dwarfs identified by \cite{Zielinski2012} we also excluded another two stars: 
TYC 3676-02387-1 ($\ALi_{NLTE}$ of 1.75) and TYC 3993-01850-1 ($\ALi_{NLTE}=1.81$) from the whole further analysis, despite  their 
high lithium abundances. According to their effective temperatures and $\log g$, these two stars are still in the subgiant phase in that they
only recently started the first dredge-up, and thus, their high lithium abundances are most likely leftover MS values.

Two more objects, as identified within PTPS as spectroscopic binaries 
with unresolved lines (TYC 3667-00550-1, TYC 4421-01996-1) and  the sixteen stars for which we only have estimates 
of the stellar parameters from  \cite{Zielinski2012} were excluded from statistical analysis for consistency as well. 
The later stars are presented in the following figures; however, they are marked as crosses colored according to their $\ALi$. 

Given that lithium depletion is not expected in a star on the  RGB before the first dredge--up (FDU) has occurred,
 it is important to determine the location of the star on the HR diagram with respect to this event. 
 For FGK stars, Li depletion due to the FDU starts at approximately $\Teff=5400$~K and ends 
at $\Teff=4500$~K \citep{CharBal2000} when the convective envelope reaches its maximum depth.
A higher Li abundance  as a remnant from the MS phase could be confused with an overabundance 
per se if the star did not experience severe FDU dilution yet.  This is particularly important
for stars in the higher mass end $\approx 1.5-3 \Msun$ and solar metallicity, since Li is not so easily 
destroyed during the MS lifetime of the star (note, however, the small number of objects we have
with solar metallicities). In solar mass stars lithium is strongly depleted during the MS lifetime 
and they are expected to leave the MS with $\ALi \sim 1$. 

Points representing the position of the RGB bottom for the presented evolutionary  tracks  were joined 
with blue striped line to define the beginning of the RGB for various masses. The  light gray area represents 
the location of the LFB on the first ascent of RGB.

In the green block in Fig. \ref{fig7} we have marked the location of the RGC region, as defined 
by 4700~K$<\Teff<5100$~K and  1.3$< \log L/L_{\odot}<$2.0  (see \citealt{Zielinski2012}).  
We can see in  Fig. \ref{fig7} that we have a substantial number of stars located in the region 
that corresponds to the first ascent on the RGB in our sample, and a handful of objects with low $\Teff$ 
and high luminosities  for which we cannot fully exclude the possibility 
that they are already evolved off the horizontal branch (HB). 

Of the 323 giants in our complete sample, 125 ($37\pm3\%$) fall into an extended RGC region of  the HR diagram as defined in \cite{Zielinski2012} 
and from the sample of 82 giants with a detected Li (group B and C) 35 ($43\pm5\%$).  
Within uncertainties, these fractions agree. 
In the LFB area of HR diagram, there are 28 ($9\pm2\%$) giants from the total sample and five with detected Li ($6\pm3\%$). 
Of 11 giants from group C, only four (36$\pm$15$\%$) fall into RGC and one (9$\pm$9$\%$) into LFB.
Again, within uncertainties, these fractions agree.
We noticed (although it is fully discussed later in the paper) that 
the Li-rich stars are not grouped in a particular region of the HR diagram in Fig. \ref{fig7}, and thus,
they do not seem to be associated with any particular evolutionary stage. 

Interestingly enough, the four most Li-rich giants in our sample form a line on HR diagram, suggesting a similar evolutionary phase; 
however,  the superposition in the HR diagram of the location of stars 
with different masses and metallicities complicates  identification of their evolutionary stage.

\subsection{Lithium abundances versus stellar parameters}

\begin{figure}
  \centering
   \includegraphics[width=0.5\textwidth]{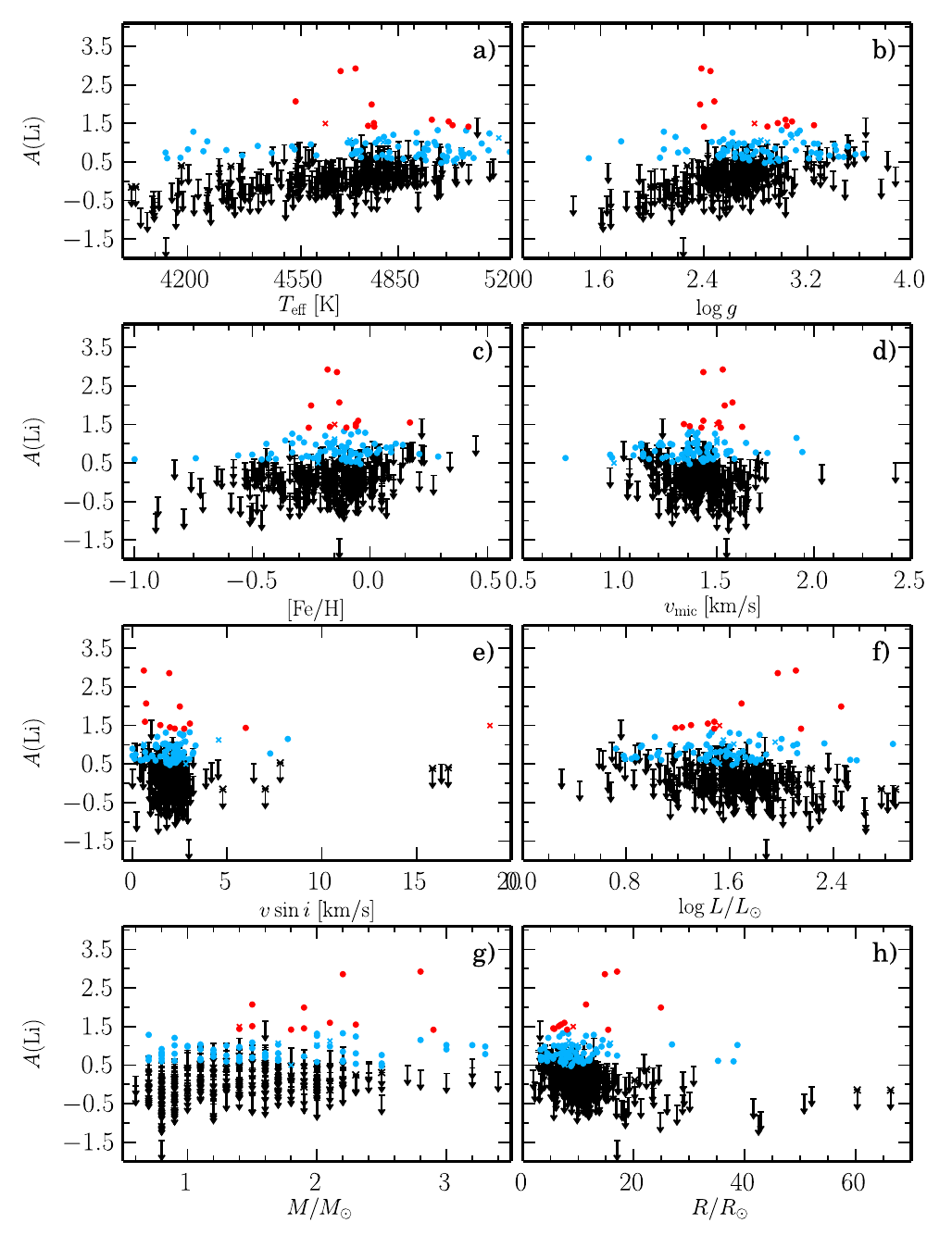}
   \caption{Lithium abundances vs. available stellar parameters.  Color coding is the same as in Fig. \ref{fig7}.}
   \label{fig8}
\end{figure}

To understand the behavior of the lithium abundances in giants, we checked for
 their possible relation with the available stellar parameters. 
Plotted in Fig. \ref{fig8} a-h  are the $\ALi$ versus spectroscopic parameters
($\Teff$, 
$\log g$, 
[Fe/H],
$\vmic$, 
$v\sin i$)
and estimated integrated parameters
(luminosities,
masses and
stellar radii). 
The points are color-coded, as in  Fig. \ref{fig7}, according to the Li content.
The mean and median values, as well as dispersions of the relevant stellar parameters obtained for the three  groups, 
are presented in Table \ref{tab6}.

%%%% 6A

The relation between $\Teff$ and $\ALi$ is explored in Figure \ref{fig8}a,
where we see an increase of $\ALi$ with effective temperature as already shown in Figure \ref{fig5}.
Giants with detected Li practically occupy the whole range of available $\Teff$, while those 
with enhanced Li abundance stay within a limited range of $\Teff>4500$~K.
The four giants with greatest $\ALi$ all have $\Teff$ in relatively narrow range between 4500 and 4800K ($4673 \pm 88$~K on average).

\begin{figure}
   \centering
   \includegraphics[width=0.5\textwidth]{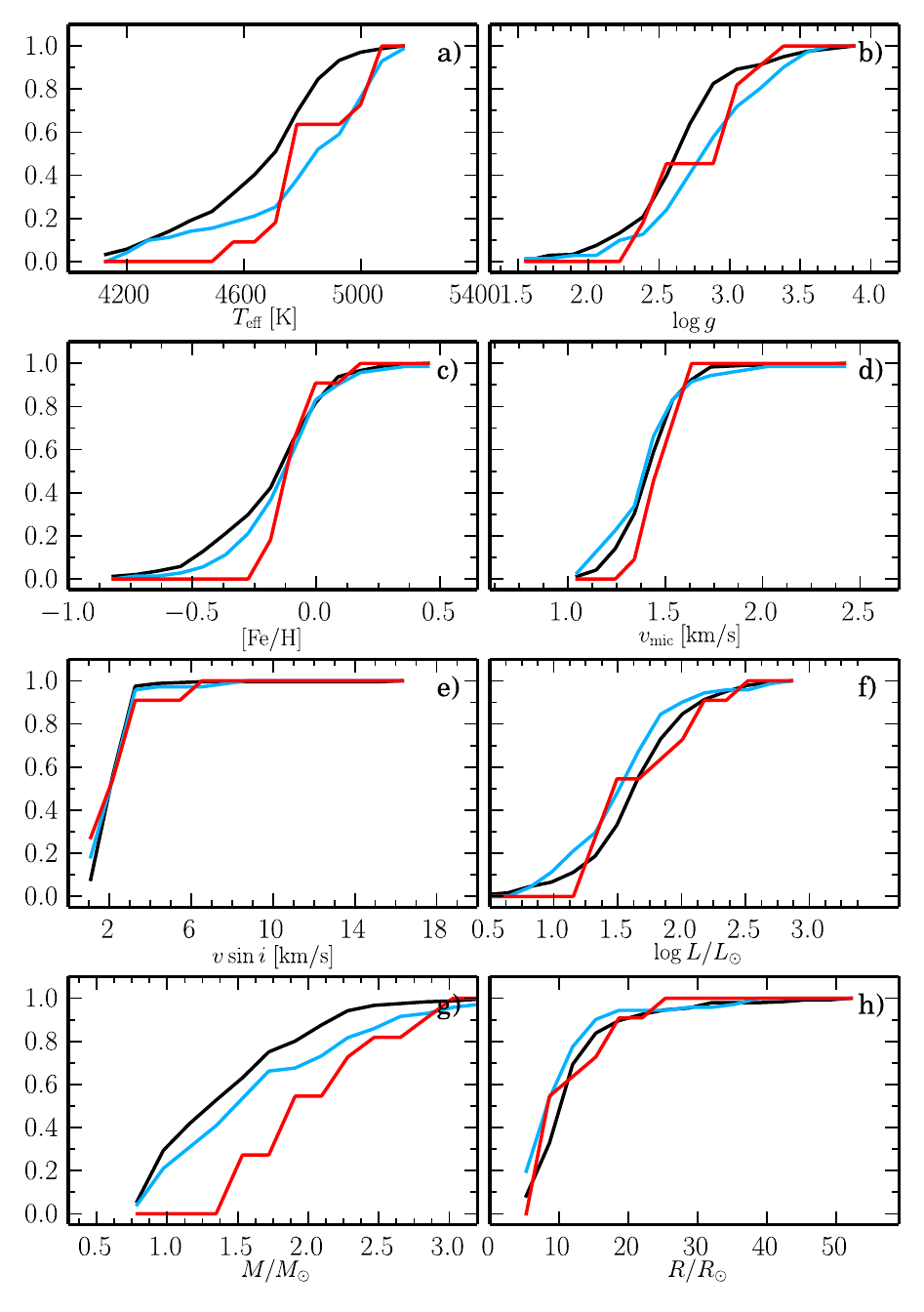}
      \caption{Cumulative distributions of available stellar parameters. Color coding is the same as in Fig. \ref{fig7}.}
         \label{fig9}
         \end{figure}

A considerable fraction
of stars occupy a narrow range of effective temperatures between 4700 to 5100~K. 
In this region, two
different evolutionary stages are expected to be co-existing: giants on their first ascent on RGB that still undergo
FDU dilution, and stars in the clump, that already burn helium steadily in their cores. The latter
group of stars should reveal lower lithium abundances compared to giants on their way to RGB tip in the absence 
of an extra-mechanism of Li production. Stars in the clump have already finished depletion due to 
the FDU, and their lithium abundances were lowered by extra mixing process, which operates 
after the RGB bump. Although separating those two groups is difficult and requires detailed spectral analysis, 
stars in the clump are expected to belong to a narrow luminosity range.  The $\Teff$  averages for  
groups A, B, and C reveal higher effective temperatures, although within uncertainties for the Li--rich stars
(see Table \ref{tab6}). 

\setcounter{table}{5}
\begin{table*}
\centering
\caption{Mean values of stellar parameters obtained for the different subsamples: (Group A) Li--poor giants (non-detections and upper limits), 
(B) giants with Li measurements, excluding Li-rich stars, 
and (C) Li-rich stars.}.
\label{tab6}
\begin{tabular}{c|ccc|ccc|ccc}
\hline
 &\multicolumn{3}{c|}{A } &\multicolumn{3}{c|}{B} &\multicolumn{3}{c|}{ C }\\
\hline
 &Median & Mean & $\sigma$& Median &Mean &$\sigma$&Median & Mean & $\sigma$\\
 \hline \hline
   $\Teff$ & 4691 & 4642 & 234 &  4836 & 4790 &269 & 4776 &  4823 & 159 \\ 
  $\log g$ & 2.64 & 2.62 & 0.40 &  2.79 & 2.81 &0.42 & 2.89 &  2.76 & 0.32 \\ 
$[$Fe/H$]$ & -0.15 & -0.19 & 0.22 &  -0.12 & -0.16 &0.20 & -0.13 &  -0.11 & 0.11 \\ 
   $\vmic$ & 1.40 & 1.41 & 0.17 &  1.38 & 1.37 &0.20 & 1.51 &  1.48 & 0.09 \\ 
 $v\sin i$ & 2.08 & 2.10 & 1.17 &  1.98 & 2.04 &1.27 & 1.99 &  2.18 & 1.45 \\ 
$\log L/L_{\odot}$& 1.62 & 1.63 & 0.42 &  1.52 & 1.52 &0.43 & 1.48 &  1.68 & 0.41 \\ 
 $M/\Msun$ & 1.30 & 1.40 & 0.55 &  1.50 & 1.62 &0.69 & 1.90 &  2.03 & 0.48 \\ 
 $R/\Rsun$ & 10.00 & 11.68 & 7.35 &  8.20 & 9.88 &6.97 & 8.00 &  11.24 & 5.87 \\ 

\hline
\end{tabular}
\end{table*}

\begin{table*}
\centering
\caption{Kolmogorov-Smirnov tests for stellar parameters. (Group A) Li-poor giants (non-detections and upper limits), (B) giants with Li measurements, excluding Li-rich stars, 
and (C) Li-rich stars. }
\label{tab7}
\begin{tabular}{c|cc|cc|cc}
\hline
 &\multicolumn{2}{c|}{A vs. B} &\multicolumn{2}{c|}{ A vs. C} &\multicolumn{2}{c|}{B vs C}\\
\hline
 & D &p& D &p& D &p\\
 \hline      \hline
       $\Teff$ & 0.37 & 0.00 & 0.39 & 0.06 &  0.27 & 0.43 \\ 
      $\log g$ & 0.26 & 0.00 & 0.37 & 0.08 &  0.31 & 0.25 \\ 
    $[$Fe/H$]$ & 0.09 & 0.73 & 0.31 & 0.23 &  0.24 & 0.58 \\ 
       $\vmic$ & 0.13 & 0.32 & 0.37 & 0.09 &  0.44 & 0.04 \\ 
    $v \sin i$ & 0.13 & 0.27 & 0.24 & 0.54 &  0.16 & 0.95 \\ 
$\log L/L_{\odot}$ & 0.17 & 0.08 & 0.22 & 0.62 &  0.24 & 0.60 \\ 
    $M/\Msun$& 0.15 & 0.14 & 0.53 & 0.00 &  0.41 & 0.06 \\ 
  $R/\Rsun$ & 0.24 & 0.00 & 0.29 & 0.28 &  0.27 & 0.45 \\ 

\hline
\end{tabular}
\end{table*}

%%% 6B
Figure  \ref{fig8}b presents a relation between  lithium abundance and the surface gravity. 
Again, giants with detected Li are present at all $\log g$ values. Stars with $\ALi$ enhanced are placed within
a range of $\log g$ between 2.4 and 3.3, while the four objects with the largest Li overabundances
seem to be located in a very narrow  range of $\log g $ and, within uncertainties, all four stars 
have identical $\log g$ of $2.42 \pm 0.05$.

Another interesting feature of Fig. \ref{fig8}b is 
a significant drop in the number of giants with detected Li  for stars with $\log g \lesssim2.4$.  
This may be, however, an artifact, which results from the $\log g$ distribution in the sample (cf. \citealt{Zielinski2012}).
Group A giants show the lowest $\log g$ on average; however, no statistically significant 
difference is found in the mean surface gravity between  groups A, B, and C.

%%% 6C
No obvious relation between $\ALi$ and [Fe/H] is present in Fig. \ref{fig8}c. 
Li seems to be detected in giants in our sample with all available [Fe/H]. 
Those with raised $\ALi$ have, however,  [Fe/H] in a range of -0.26 to 0.17. 
The four most Li abundant giants in our sample show practically identical metallicities of $-0.17 \pm0.05$.
Giants with detected Li (group B) and Li--rich giants (group C) seem to have higher metallicities on average 
(see Table \ref{tab6}), but this might just reflect poor statistics.

%%% 6D
In Fig. \ref{fig8}d, we see that giants with detected Li present  microturbulence velocities in the whole range available in our sample. 
The Li-abundant ones, group C, occupy a tighter range of  $\vmic=1.3-1.6$, and the four giants 
with largest $\ALi$ have $\vmic=1.52 \pm 0.06 \kms$, which is identical within uncertainties.
No meaningful differences exist between mean values in groups A, B, and C.

%%% 6E
 The rotational velocities $v\sin i$ (obtained during SME analysis) versus $\ALi$ are presented in Figure  \ref{fig8}e. 
 Fast rotators in the RGB are defined as stars with $v\sin i \gtrsim 10 \kms$ \citep{Fekel1997}. 
 Most of our stars have rotation velocities expected during the RGB as a consequence 
 of conservation of angular momentum; that is, $v\sin i \approx 2-3 \kms$ \citep{Grey1989, deMedeiros1999, Fekel1997}. 
 The possible relation of stellar rotation with the overabundance of Li has been recently explored by \cite{Carlberg2012}, 
 where the authors found that rapid rotators are on average enriched in lithium when compared to the slow rotators. 
 All four stars that show the largest amount of lithium have $v\sin i <5 \kms$.  Moreover, the average rotation 
 velocity of all the Li-detected stars are very similar.
The only exception is TYC~0405-01700-1, which is plotted in Fig. \ref{fig8}e with a red cross. 
With uncertain stellar parameters, this star shows broad line profiles that may by partly 
attributed to rotation but  it shows RV variations of SB2 type at the same time. Its actual status 
can be resolved with a more detailed spectroscopic analysis, which is beyond the scope of this paper.

%%% 6F
With regard to luminosity, Fig. \ref{fig8}f,  we note that  the giants with Li detection are present among 
stars in a wide range of luminosities  $\log L/L_{\odot}> 0.7$, which practically in all our samples.
The giants with enhanced $\ALi$ all have $\log L/L_{\odot}> 1.2$,
and the four most Li abundant ones have $\log L/L_{\odot}$ in 1.6-2.4 range ($2.06 \pm 0.28$ on average). 

%%% 6 G
Giants with Li detection also practically cover the whole range of stellar masses as illustrated in 
Fig. \ref{fig8}g.  The higher the Li content, the higher the stellar mass, and Li enhanced giants 
all show masses larger than $1.4 \Msun$. The most  Li abundant giants belong to most 
massive ones in our sample. The four most Li abundant giants have $M/\Msun=2.1 \pm 0.5$ on average, and
the two  most Li rich  have masses of  $2-3 M_{\odot}$, which is higher above average. 
The mean values in groups A, B, and C are presented in Table \ref{tab6}. 
They  support a $\ALi$ - stellar mass relation \citep{Mallik1999}, although mean values are
associated with large uncertainties.

%%% 6H
In Fig. \ref{fig8}h, we see that giants with Li detection or overabundance tend to have moderate radii, up to about $39 \Rsun$. 
This is especially true for the four giants with the largest Li abundance and with radii ranging from 11 to 25 solar radii ($17\pm 5 \Rsun$ on average).
The lack of Li detection among the giants with the largest  radii, however, may very well  only reflect the small number of stars from our sample in that range.
The average values do not show significant differences among stars for various groups.

In general, we note that the differences in stellar parameters between groups A, B, and C are not obvious. 
However, interestingly enough, the four  most Li abundant giants in our sample 
(TYC~0684-00553-1, TYC~3105-00152-1, TYC~3304-00090-1, and TYC~3917-01107-1) 
seem to have very similar stellar parameters, are not fast rotators, and have surprisingly similar $\log g$, metallicity and microturbulence velocity. 
The largest discrepancies  are present in the least 
constrained ones: masses, radii and luminosities. 
This is especially true for the two giants with the largest $\ALi$, TYC~0684-00553-1 and TYC~3105-00152-1,
with almost identical metallicities, luminosities, masses, and radii.
They also have masses at the borderline among solar, intermediate, and moderate radii. 
With their location on HR diagram (Fig. \ref{fig7}), these results suggest some specific evolutionary phase.
 
Kolmogorov--Smirnov (K-S) tests on the cumulative distribution function (CDF) 
of the three groups of stars that are defined according to their  lithium content 
and eight stellar parameters are shown in Fig. \ref{fig9} and summarized in Table  \ref{tab7}, 
where the D parameter (K-S statistic)
and the p-value (probability that the two samples are driven from the same distribution) are given
for each comparison between pairs. 

The cumulative distributions of stellar parameters generally confirm our coarse 
analysis of  Figure \ref{fig8}. 
The most striking, statistically significant differences between groups A and B are present in distributions of effective temperatures, 
$\log g$, and radii. Stars with detected Li are 148 K ($<$1$\sigma$) hotter on average, have a larger $\log g$ 
(by 0.19, again $<1\sigma$), and smaller radii ($1.8 \Rsun$, again $<1\sigma$).

One can see from Fig. \ref{fig9} and Tables  \ref{tab6} and \ref{tab7} 
that the largest, statistically significant  
differences between samples A and C in CDFs exists  in stellar masses. Li-rich giants seem to 
favor higher stellar masses (by $0.6 \Msun$ or about $1 \sigma$).  
Giants with Li detection (group B) are also less massive than those with Li overabundance (group C).

A less noticeable difference also exists in effective temperature distributions, stars from group C 
are 181~K (about $1 \sigma$) hotter.  Groups A and C  also show less prominent discrepancies in $\log g$ and $\vmic$.
The CDF for surface gravity shows a gap for Li-rich giants in the range of $\log g \approx 2.5-3$ 
that may be connected to the substantial drop of Li abundances present in Fig. \ref{fig8} 
around this gravity range, but it can also be a result of a poor statistics in this region. 

The largest differences between groups B and C are distributions of $\vmic$ and stellar masses. 
In the case of stellar masses, group B giants show distributions more like group A, while 
$\vmic$ distributions in groups A and B are indistinguishable. The Li-rich giants have  
the largest $\vmic$ on average, which is larger by 0.11 ($\sim1\sigma$) than giants with detected Li only.

The two-sample K-S test confirms that all three distributions are the same from a statistical
 point of view, in the case of  metallicity, $v \sin i$, and luminosity.
 
\begin{figure}
   \centering
   \includegraphics[width=0.45\textwidth]{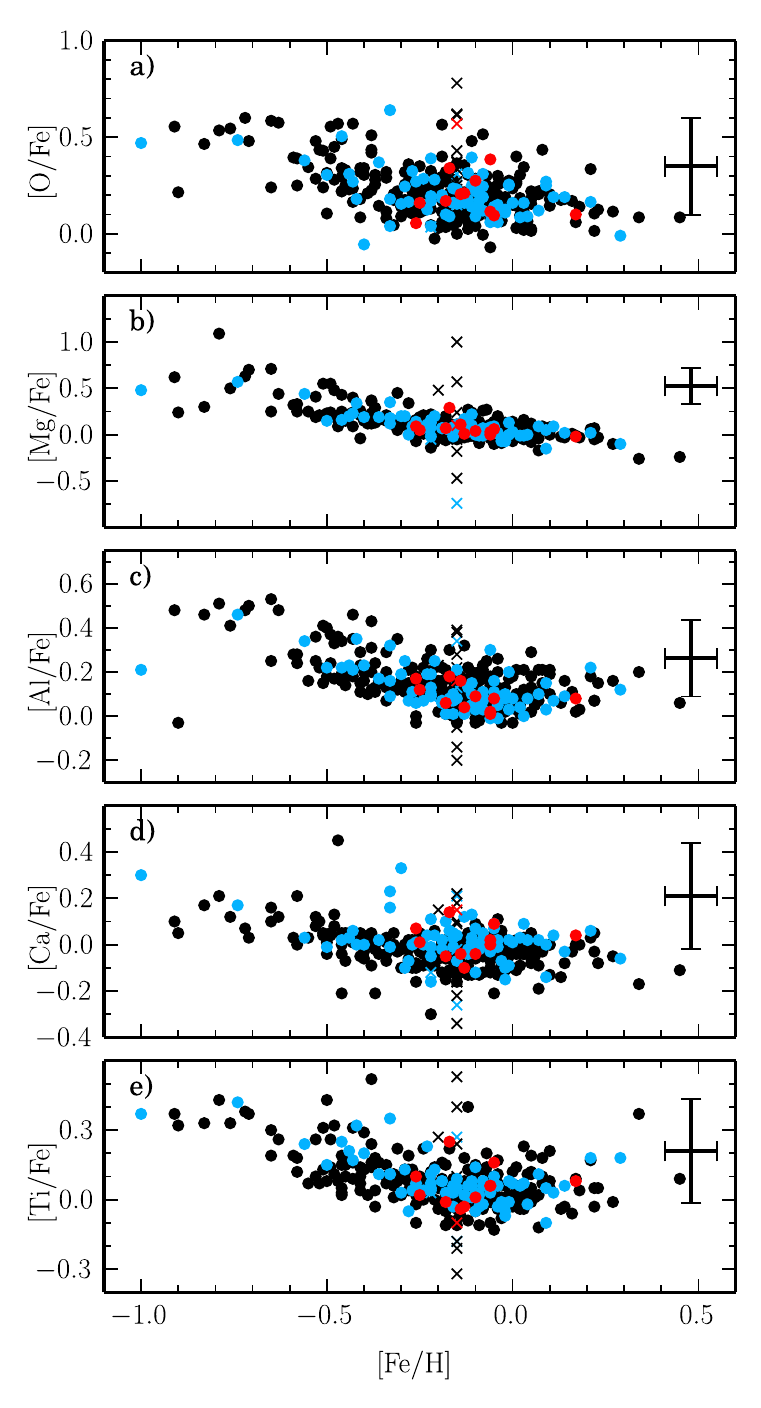}
      \caption{ [$\alpha$/Fe]  abundances versus metallicity. Color and points coding are the same as in Fig. \ref{fig7}.}
         \label{fig10}
         \end{figure}

\begin{figure}
   \centering
   \includegraphics[width=0.45\textwidth]{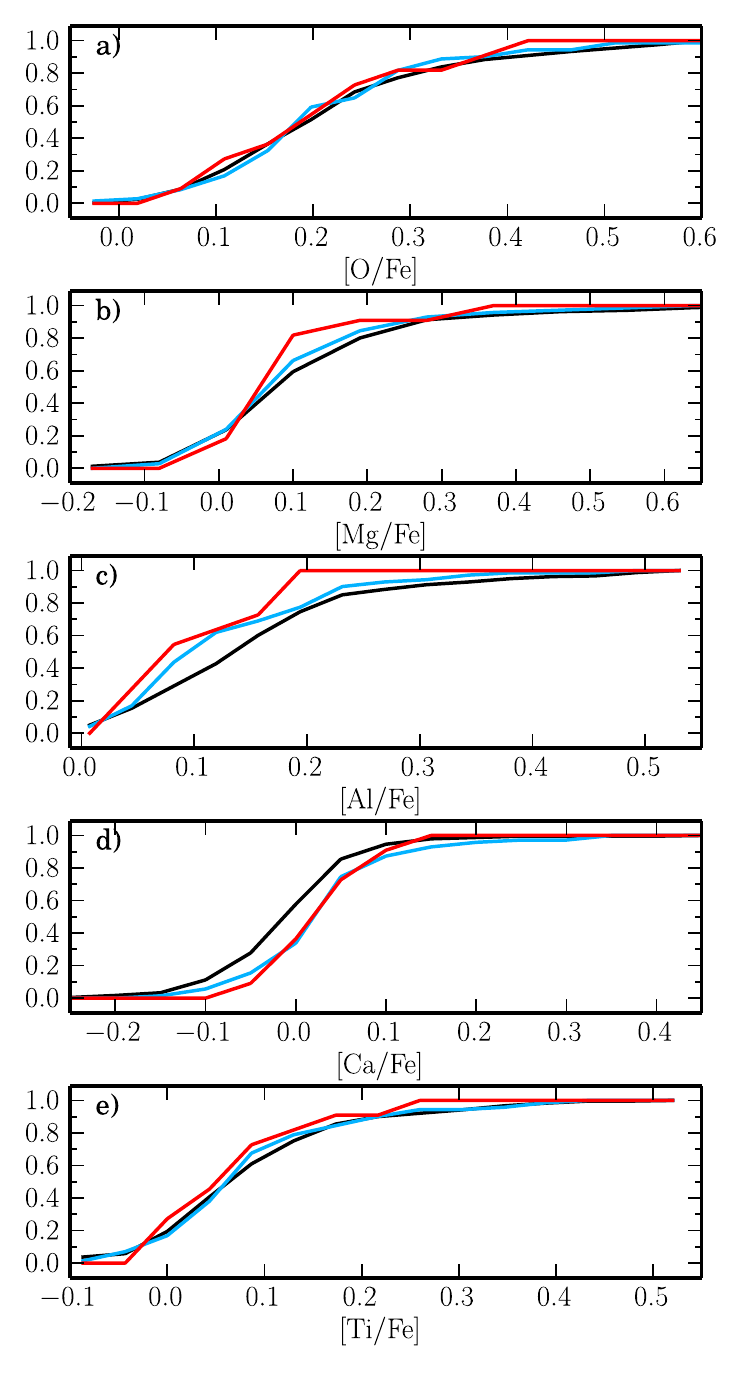}
      \caption{Cumulative distributions of [X/Fe] values. Color coding is the same as in Fig. \ref{fig7}.}
         \label{fig11}
         \end{figure}

\subsection{ Lithium vs. $\alpha$-elements abundances}

We  searched for anomalies in terms of the composition of the $\alpha$-elements 
for the stars belonging to the three groups, defined according to their Li content. 
These elements are not produced in this mass range (oxygen the exception) and are certainly 
not expected to be processed during the RGB evolution of a star. 

It has been suggested that the [$\alpha$/Fe] trend with [Fe/H] is due to the time delay between  
SN II, which produces alpha elements and iron-peak elements \citep{Arnett1978, WoosleyWeaver1995}, 
and SN Ia, which yields mostly iron-peak with little alpha element production.  
To check unusual chemical compositions associated to SN II explosions in the sample 
and whether there is any trace of enhanced  $\ALi$ to be associated with such a deviation,
we investigated abundance trends with metallicity for several $\alpha$ elements (O, Mg, Ti, Ca) 
and for aluminum, sometimes defined as mild-$\alpha$ element. 
The [X/Fe] abundances, as a function of the [Fe/H]  are shown in Figure \ref{fig10}.
The mean and median values of [X/Fe] abundances, as well as dispersions, are presented 
for all three groups in Table \ref{tab8}. The CDF are presented in Figure \ref{fig11}, 
and the results of K-S tests are listed in Table \ref{tab9}.

\begin{table*}
\centering
\caption{Mean values of abundances obtained for the different subsamples. (Group A ) Li--poor giants, 
(B) giants with Li measurements, excluding Li-rich stars, and (C) Li-rich stars.}.
\label{tab8}
\begin{tabular}{c|ccc|ccc|ccc}
\hline
 &\multicolumn{3}{c|}{A } &\multicolumn{3}{c|}{B} &\multicolumn{3}{c|}{ C }\\
\hline
 &Median & Mean & $\sigma$& Median &Mean &$\sigma$&Median & Mean & $\sigma$\\
 \hline \hline

 $[$O/Fe$]$ & 0.19 & 0.21 & 0.13 &  0.18 & 0.21 & 0.12 & 0.17 &  0.19 & 0.10 \\
$[$Mg/Fe$]$ & 0.08 & 0.11 & 0.16 &  0.07 & 0.09 & 0.12 & 0.05 &  0.07 & 0.08 \\
$[$Al/Fe$]$ & 0.13 & 0.15 & 0.11 &  0.09 & 0.12 & 0.09 & 0.08 &  0.09 & 0.06 \\
$[$Ca/Fe$]$ &-0.02 &-0.02 & 0.08 &  0.01 & 0.02 & 0.09 & 0.01 &  0.01 & 0.07 \\
$[$Ti/Fe$]$ & 0.06 & 0.08 & 0.11 &  0.06 & 0.08 & 0.10 & 0.06 &  0.06 & 0.08 \\
\hline   

\end{tabular}
\end{table*}

\begin{table*}
\centering
\caption{Kolmogorov--Smirnov tests for abundances. (Group A) Li-poor giants, (B) giants with Li measurements, excluding Li-rich stars, 
and (C) Li-rich stars.}
\label{tab9}
\begin{tabular}{c|cc|cc|cc}
\hline
 &\multicolumn{2}{c|}{A vs. B} &\multicolumn{2}{c|}{ A vs. C} &\multicolumn{2}{c|}{B vs C} \\
\hline
 & D &p& D &p& D &p\\
 \hline \hline

$[$O/Fe$]$ & 0.10 & 0.62 & 0.27 & 0.37 &  0.21 & 0.74 \\ 
$[$Mg/Fe$]$ & 0.08 & 0.83 & 0.24 & 0.51 &  0.20 & 0.81 \\ 
$[$Al/Fe$]$ & 0.13 & 0.29 & 0.29 & 0.30 &  0.21 & 0.74 \\ 
$[$Ca/Fe$]$ & 0.10 & 0.58 & 0.21 & 0.68 &  0.22 & 0.69 \\ 
$[$Ti/Fe$]$ & 0.10 & 0.67 & 0.24 & 0.54 &  0.22 & 0.69 \\
      \hline      
\end{tabular}
\end{table*}

As it was in the case of stellar parameters, we can see that giants with detected Li are present in whole 
range of abundances covered with our sample. Li-rich giants occupy generally a more narrow range, reflecting their metallicities.
We see for all elements  the expected trend for a lower [X/Fe] ratio as the metallicity increases. 
Around solar metallicities, the relation becomes more flat, which is attributed
 to the possible onset of SN Ia (see \citealt{McWilliam1997}). Only few stars stand out from the relation
 and are those with large uncertainties in their  abundances determination or stars rejected from statistical 
 analysis due to uncertain stellar parameters or identified as spectroscopic binaries. 
Lithium rich stars follow the same overall trend with [Fe/H] as all the $\alpha$ elements analyzed.

The K-S tests confirm that there are not statistically significant differences between $\alpha$-elements 
abundances CDF for groups A, B, and C.
In the case of aluminum and magnesium, we note a systematic effect:
the giants with Li detection or overabundance show lower values 
of [Al/Fe] and  [Mg/Fe], however, at the level well below $1\sigma$. 
We find no statistically significant trace of $\alpha$-elements enhancement in giants with detectable Li content.
The only outlier is TYC 0405-1700-1, which presents slight O enhancement.

\subsection{Li overabundance and signatures of mass loss}

If planet engulfment was the cause of lithium abundance anomalies,
 then it would be reasonable to expect a relation of lithium abundance and potential mass ejections 
 from the star (see e.g. \citealt{delaRezaDrake1997, SiessLivio1999,Lebzelter2012}) 
 that might show as infrared excesses. We have collected data from 
 WISE \citep{WISE2010}, IRAS \citep{Neugebauer1984}, and 2MASS catalogs 
 for the stars in our sample in  search for infrared excesses. A total of 317  stars from the 323 sample, 
 16 giants with uncertain parameters, and two spectroscopic binaries were 
 identified in the WISE catalog with  only four stars falling outside of the 1'' search radius 
 and 83 matches within a 25'' search radius were found in the IRAS catalog
 (queried in both the Point Source Catalog and Faint Source Catalog).

\begin{figure}
   \centering
   \includegraphics[width=0.5\textwidth]{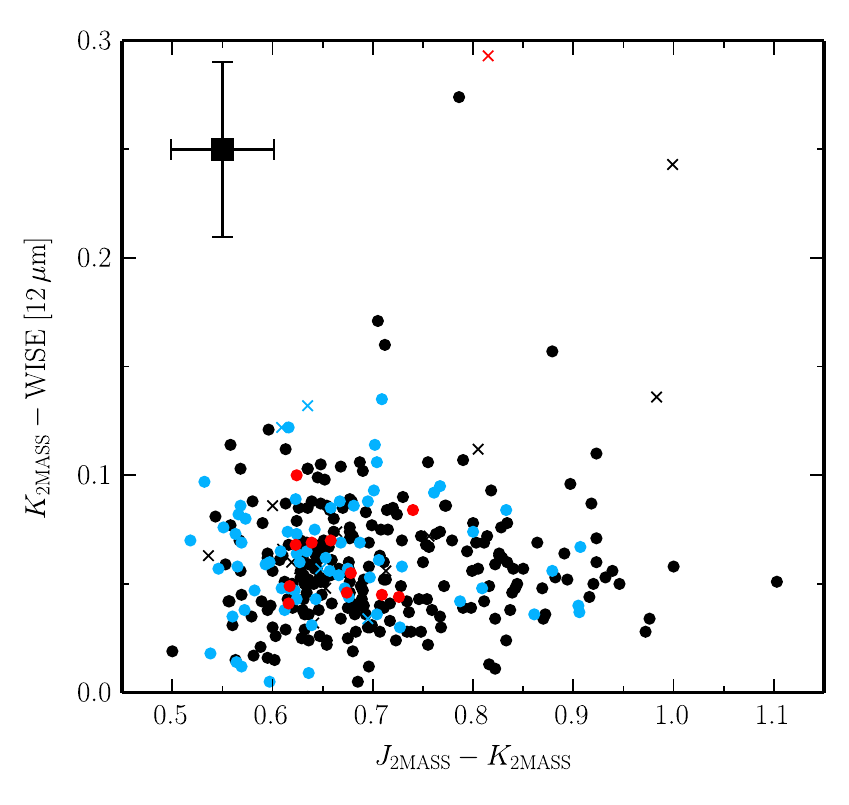}
      \caption{ K-WISE[12$\mu$m] color-- versus the J-K 2MASS color diagram. The 
outlier in the K-WISE[12$\mu$m] color is TYC 0405-01700-1. Color coding is the same as in Fig. \ref{fig7}.} 
         \label{fig12}
         \end{figure}

In Fig. \ref{fig12}, we present the K-WISE[12$\mu$m] color versus the J-K 2MASS 
color diagram of our stars, based on the color selection presented by \cite{Lebzelter2012}, 
who investigated the possibility of enhanced mass-loss in lithium rich stars. Only few stars,
 including one Li-rich star (TYC 0405-01700), are outliers from the general flat relation K-WISE[12$\mu$m] 
=0.06. However, with K-WISE[12$\mu$m] $<$1 they do not appear to be objects with  substantial mass loss taking place \citep{Lebzelter2012}.

\begin{figure}
   \centering
   \includegraphics[width=0.5\textwidth]{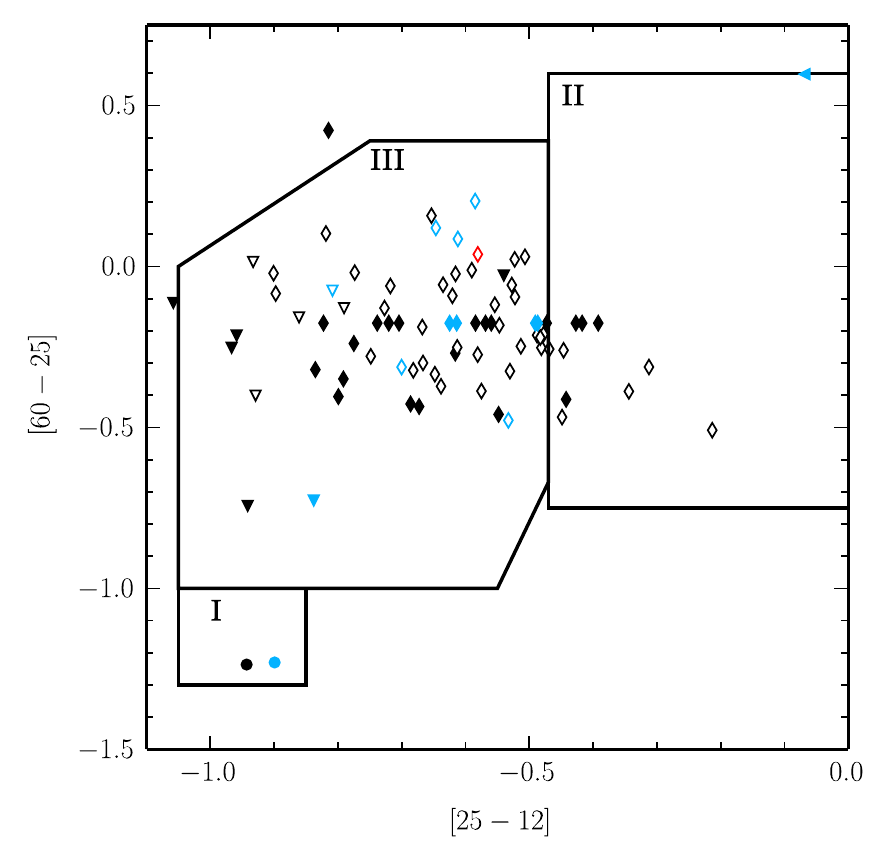}
      \caption{The Li--cycle diagram  according to \cite{delaRezaDrake1997}. Filled and open symbols represent different IRAS catalogs: 
      PSC - Point Source Catalog, FSC - Faint Source Catalog, respectively. Color coding is the same as in Fig. \ref{fig7}.
      Symbols stand for IRAS fluxes quality: circles - both colors have good quality, triangles pointing left - upper limit for [25-12], 
      triangles pointing down - upper limit for [60-25], diamonds - upper limits for both colors.}
               \label{fig13}
         \end{figure}

Figure \ref{fig13}  shows an IRAS two color-color diagram with three regions defined by \cite{delaRezaDrake1997} 
based on mass loss rates and the IR excess consequence of it. In the region labeled as I no Li--rich nor fast rotating 
giants are expected as, according to \cite{delaRezaDrake1997}, in this region only ''normal'' K giants, i.e. stars with 
low lithium abundances and no signatures of circumstellar shell should be present. Region II delimitates a color-color
 region with infrared excesses.  If lithium production was associated to a planet engulfment episode in the form of 
 a mass loss increase then the star should move to region II where infrared excess is observed. When this episode 
 ends, the star should move to region III, where a small infrared excess is still present, and finally 
 will head back to region I closing the cycle.

Region I is poorly populated - only two stars are located
 in here - TYC 3226-02285-1 and TYC 3930-01790-1, one with detected Li. These are also the only two stars 
 with both IRAS colors available. For all remaining stars only upper limits in one or both IRAS colors are available.
In the most interesting region II we find 10 stars, all with upper limits of Li abundances.
 The vast majority of the stars identified  in the IRAS catalog, 69 out of 83 ($83\pm4\%$), are located in region III. 
 In that region  we identified super-Li star TYC 3917-01107-1, 11 stars with detected Li (group B) and 58 from group A. 
 Only $16\pm5\%$ of stars in that region belong to group B or C.

What Fig. \ref{fig13} seems to imply is that most of giants identified in IRAS catalog have moderate infrared excesses. 
These excesses, if triggered by an engulfment process, would suggest that it has happened recently,
 and we observe stars before their envelopes became diluted. 
 It  seems more reasonable  to assume that if the nature 
 of the excess is real, then it is most likely associated with RGB mass loss than due to an extra mechanisms. 

Another empirical way of investigating the mass loss phenomena in our sample  is through the Na D lines profiles.
In cases of intense mass loss, an additional component in the sodium lines should be observed on the blue side of the 
Na structure and should be separated from the stellar line from several up to a dozen $\kms$ \citep{Reddy2002}.
The problem with the Na D line features in our spectra is that the data reduction process needs to be done 
extremely  carefully due to possible contamination of the stellar spectra by emission structures in the flat-field lamp.
This spectral range requires  also removal of the iodine lines (which is 
a part of RV code for line bisector calculations) but  introduces  extra noise.
We have performed this exercise but only for the stars that belong to the Li-rich sample (group C). 
Complicated structures around Na D lines are present for some stars. 
For example, TYC 3663-01966-1 presents  additional components on the blue side of spectra. 
However, we found it difficult  to identify their origin and assign it to circumstellar gas instead of to the interstellar medium. 

Intensive mass outflows might also have the effect of changing the shape of  observed  spectral lines, a phenomena 
that should be observable in a few years span. Since PTPS data has been collected since 2004,  we checked 
for differences in the line shape of selected stars. We analyzed the sample of Li-rich 
stars, and we did not find any significant differences in the line shapes  that could be attributed to mass loss.

\section{Analysis of the individual stars that present Li overabundances\label{stars}}
In this section, we present detailed, individual comments of the 12 giant stars that show high Li abundances (group C and TYC 0405-01700-1). 
Their parameters are listed in Table \ref{tab10}.

\begin{table*}
\centering
\caption{Stellar parameters for the 11most Li abundant giants in our sample and TYC 0405-01700.}
\label{tab10}
\begin{tabular}{c |cr ccc crr lcc llc}
\hline 
TYC  & $\Teff $&Fe/H  & $\ALi$&$v \sin i$ & $\log L/L_{\odot}$ & $M$                  &$R$& RV  status&infrared &comment\\
                         &[K]        &             &NLTE               &$[\kms]$ &                  &[$\Msun$]  &[$\Rsun$]                        &                  &       photometry  &              \\
\hline \hline
0684-00553-1 & 4719 &  -0.18   & 2.92 & 0.6         & {2.11} & {2.8} &  17.0 &RV var 			&-&\\
3105-00152-1 & 4673 & -0.14    & 2.86 & 2.0          & 1.97 	  & {2.2} & 14.8 & RV var			 &-&\\
3304-00090-1 & 4534 &-0.13     & 2.07 & 0.7          & 1.69 	  & 1.5 	& 11.4 &planet		 & &\citealt{Adamow2012}\\
3917-01107-1 & 4769 & -0.25    &  1.99& 2.5         & {2.46} & 1.9 	& 24.9&planet		 &IRAS, III & Niedzielski et al. - in prep.\\

\hline
0435-03332-1 & 4955 &  -0.05   & 1.60 & 0.7          & 1.48       & {2.1} &  7.5 &single 			&-	&\\
1058-02865-1 & 4758 &  -0.17   & 1.44& {6.0}   & 1.18 	  & 1.4 	& 5.7 &single		 	&-&\\
3300-00133-1 & 5007 &  0.17    &  1.55 & 3.0          & 1.43         &2.3      &  6.9&single                                & -&\\
3314-01371-1 & 5019 &-0.06     & 1.45 & 2.0          & 1.23 	  & 1.9 	& 5.5&SB1			 &-&\\
3318-01333-1 & 4776 &-0.06     &  1.51& 1.5          & 1.30 	  & 1.5 	& 6.5&RV var 			&-& \\
3663-01966-1 & 5068 & -0.26    & 1.41 & 2.7          & {2.15} & {2.9} & 15.4 &single 			&-&\\
3930-00681-1 & 4777 &-0.10     & 1.42 & 2.2          &  1.48	 & 1.8 	&  8.0 &single		  &-&\\
\hline
\hline
0405-01700-1 & (4626)& (-0.15)  & (1.50) &{(18.9)} & (1.52)      & (1.4) 	&  (9.1) & SB2 			 &{ WISE} & est. stellar params.\\
\hline
\end{tabular}
\end{table*}

\subsection{Super-Li giants}

The object TYC~0684-00553-1 with  $\ALi>3.3$ is the star with the largest Li content identified in this work.
Its Li abundance is close to the meteoritic value, so it is clear that a Li-enhancement process 
must have been operating in this star. With the exception of the high $\ALi$, 
we found no signs of any enrichment processes: this star shows normal 
abundances of other elements studied here, slow rotation, and no presence of IR excess in WISE photometry 
(no matching identification was found in the IRAS catalog).
With a mass of $2.8 \Msun$, this star does not undergo a luminosity bump on the RGB, and regarding its location in the HR diagram, 
it might be in the clump or post clump phase. The HET/HRS radial velocities collected so far for TYC 0684-00553-1
reveal chaotic changes with an amplitude of $\sim100 \ms$, which exclude a stellar companion, 
but give no constraints on planetary companion, stellar activity, or pulsations yet.

The star TYC~3105-00152-1 is very similar  to TYC~0684-00553-1
 in terms of effective temperature ($\Teff=4673$~K), $\log g$, and metallicity. 
It is also the second giant star with a Li abundance close to meteoritic value 
identified in this work. With a mass of $2.2 \Msun$  it is expected, unlike  TYC 0684-00553-1, to experience the 
luminosity bump on the RGB. For $2.2\Msun$ stars with solar metallicity, the LFB 
occurs at $\Teff \sim 4400$~K  \citep{CharBal2000}. Although our luminosity and masses are uncertain,  
the uncertainty in $\Teff$ is very small (15~K, determined by \citealt{Zielinski2012}), 
and the star does not seem to be associated with the LFB. Precise HET/HRS RV for this star reveal 
an amplitude of $\sim100 \ms$, which, again, seem to exclude a stellar companion 
but are not numerous enough for a detailed analysis yet.

The object TYC~3304-00090-1 = BD+48 740 with $\ALi=2.2$ and a radial velocity amplitude of 70~m~s$^{-1}$
was analyzed in detail by \cite{Adamow2012}. This is the third most Li-rich giant in our sample, which is 
an  interesting star hosting a planet. 
The Li abundance analysis has been 
extended with new epochs of HET/HRS observations, so we present an
updated lithium abundance of $\ALi=2.07$ here. In the HR diagram this star is located near LFB, however 
its $\log L/L_{\odot}$ is uncertain.
The preliminary determined eccentric 
orbit of the observed planet with the Li enrichment 
suggest a recent engulfment of a putative second planet as a very plausible scenario.
However, a slow rotation rate and the absence of IR excess  do not support this conclusion.
An extensive analysis of this system is in preparation.

The giant TYC 3917-01107-1, a  $1.9 \Msun$ star is  another example of a planetary system 
around a Li-rich giant  (Niedzielski et al.  - in prep.). This is the fourth most Li-rich star in our sample.
In spite  of large uncertainties in  luminosity  ($\log L/L_{\odot}=2.46 \pm 0.6$), the star seems to be  located above the clump region.
 With the effective temperature of  $\Teff=4769$~K, it is unlikely that it is currently undergoing  LFB
(for $2\Msun$, LFB occurs at $\Teff \sim 4400$~K). Thus, lithium enrichment does not seem 
to be an effect of Li production during LFB. 
TYC~3917-01107-1 is the only super-Li giant in our sample with IRAS data that point to a trace of IR excess,
which may be a result of mass loss and makes engulfment scenario more likely.

\subsection{Li-rich giants}

The star TYC 0435-03332-1 is  located on the HR diagram in the RGC region;
however we cannot rule out the possibility that it is a giant in the early RGB 
with ongoing Li dilution caused by the FDU.
Based on stellar evolutionary tracks and observed Li abundances for MS and subgiant stars \citep{Ramirez2012, Mallik1999}, 
we can assume that this star left MS with $\ALi \sim 2 - 2.5$. The current Li level suggests a dilution by up to 1~dex.
Only a few epochs of HET/HRS RVs, are available for this star as it was classified in PTPS as a single.

The object TYC 1058-02865-1 has a low luminosity ($\log L/L_{\odot}=1.18$), solar-mass ($1.4 \Msun$), $\Teff=4758$ and $\ALi=1.4$. 
This star should have left the MS with $\ALi\sim 2$, and it is most likely still undergoing the FDU process. 
In PTPS, after four epochs of HET/HRS RVs that reveals an amplitude below 50 $\ms$, it was classified as a single star. 
Compared to other stars in the group of Li-rich objects, it reveals a relatively fast rotation velocity (6~$\kms$). 

The star TYC 3300-00133-1 has moderate LTE Li abundance of  $\ALi =1.25$
but  becomes Li-rich after adding the non--LTE correction of 0.18.  It is located in RGC region, 
but it is also quite massive  ($M=2.3\Msun$), so it cannot be excluded that this star is still undergoing 
the FDU  dilution. In PTPS, it was classified as a single.

The giant TYC 3314-01371-1 is not an  RGC object because of its low luminosity and mass of $1.9\Msun$;
thus lithium depletion in the FDU might be still taking place. 
After 15 epochs of HET/HRS, this star has revealed an RV amplitude above $7 \kms$. 
A preliminary Keplerian fit (Table \ref{binaries}) shows that we are dealing 
with a subsolar minimum mass companion of $0.24 \Msun$ in a long period, 845 days, with
eccentric ($e=0.34$) orbit. No trace of additional line system was 
found in our CCF analysis, so the star was classified as SB1 in PTPS.
In addition to quite high lithium abundance, we observed a slightly 
elevated oxygen abundance, suggesting external enrichment, 
and no clues for mass transfer. 

The object TYC 3318-01333-1 is another  $1.5\Msun$ star, which is not in the RGC
because of its low luminosity ($\log L/L{_{\odot}}=1.3$). 
Therefore, it is probably depleting lithium in FDU.
Collected HET/HRS RVs show low amplitude, non-periodic variations.

The star TYC 3663-01966-1 is the most massive star ($2.9\Msun$) in our Li--rich sample. 
With $\log L/L_{\odot}$ of 2.15, it is located above RGC, which is close to the early AGB. 
The $\ALi_{NLTE}$ of 1.41 and the observation, that stars of that mass leave the MS with $\ALi\sim 3$, 
suggest that a significant depletion process has already occurred.
Another interesting property of this object is its Na profile, which reveals  
a complicated structure with four non-stellar components on the blue side of the stellar lines. 
They possibly originate in a circumstellar environment and may be associated with intensive mass loss phenomenon. 
However, considering that observed lithium abundance is probably due to FDU dilution, 
 intensive mass loss should not be address to Li enhancement process. 
Radial velocities collected within  PTPS show chaotic, low amplitude variations that cannot be
 clearly associated with a companions or periodic stellar activity, and the star was assumed to be single.

The giant TYC 3930-00681-1 is located in the RGC region ($\Teff=4777\; \mathrm{K}, \log L/L_{\odot}=1.48$). 
With a mass of $1.8 \Msun$, this star should have left 
the MS with $\ALi$ close to 2. It cannot be excluded that the observed $\ALi=1.42$ 
is due to the FDU depletion that has not been completed yet. 
After a few epochs of precise HET/HRS RV, this star was assumed as single.

\begin{table}
\centering
\caption{Preliminary orbital parameters of Li-abundant giants in binary systems.}
\label{binaries}
\begin{tabular}{lll}
\hline
 Star					               & TYC 0405-01700-1 	      	  & TYC 3314-01371-1		    \\ \hline \hline
 $P\thinspace[\rm{days}]$                  & $31.753 \pm 0.0009$     & $844.6 \pm 0.6$       \\
 $T_0\thinspace[\rm{MJD}]$              & $53186.6 \pm 0.45$        & $54498.90 \pm 0.82$   \\
 $K\thinspace[\kms]$                          & $28.88 \pm 0.18$           & $3.686 \pm 0.013$      \\
 $e$                                                    & $0.040 \pm 0.004$         & $0.3392 \pm 0.0034$    \\
 $\omega\thinspace[\rm{deg}]$          & $315 \pm 5$                   & $58.7 \pm 0.5$        \\
 $m_2\sin i\thinspace[\Msun]$           & $0.53 \pm 0.08$              & $0.235 \pm 0.022$      \\
 $a\thinspace[\rm{AU}]$ 		        & $0.220 \pm 0.016$          & $2.17 \pm 0.08$        \\
 $V_{0}\thinspace[\kms]$        	        & $-10.59 \pm 0.12$           & $0.758 \pm 0.009$    \\
 $\sqrt{\chi_\nu^2}$             	        & $0.80$                             & $0.92$                 \\
 $\sigma_{\rm{RV}}\thinspace[\ms]$  & $84.72$     			 & $21.80$        \\
 $N_{\rm{obs}}$                                  & $7$                                 & $15$                  \\
 \hline
\end{tabular}
\end{table}

\subsection{TYC 0405-01700-1}{\label{4051700}

This star belongs to the group of 16 giants for which only approximate stellar parameters
 are available in \cite{Zielinski2012}, but it reveals 
several interesting characteristics. 
Seven epochs of HET/HRS RV measurements available for this star  have shown an amplitude of over $40 \kms$; 
therefore, it was classified in the PTPS as a binary.
Based on preliminary 
Keplerian orbital parameters  (Table \ref{binaries}), we estimated an orbital period  of 31.7 days, 
which agrees with \cite{Kiraga2012}, who classify the star as RS CVn.
We also estimate the minimum mass 
of the companion to $0.53\Msun$  
and postulate that the less massive companion is a more evolved star. 
 
The CCF analysis shows a distortion in the spectral lines, 
which might be due to an unresolved, second set of lines from other object;
thus, all parameters coming directly from spectral analysis should be considered very uncertain.
This star seems to be a fast rotator with $v \sin i =19 \kms$. 
It shows the reddest color in WISE and 2MASS photometry, and besides 
the high Li abundance, it also shows oxygen enhancement for its adopted metallicity. 
It is interesting to note that TYC 0405-01700-1 was identified as a 2MASS 
object that coincides with a Rosat X-ray source \citep{HaakonsenRutledge2009}, and it might exhibit coronal activity. 
In such a case, the observed Li enhancement might stem from mass transfer from an evolved stellar
companion, which would explain the abnormal chemical composition 
of TYC~0405-01700-1, its fast rotation, and its mild IR excess. 
Low eccentricity of the system
and a very short period, resulting in  orbital separation of only $\sim 5$ stellar radii seem 
to agree with this scenario. We note, however, that  
the current secondary mass is only $0.75\Msun$ if we assume average value of $\sin i$,
and the system must have lost/transferred at least  $0.65\Msun$.
On the HR diagram, this star is located near the luminosity function bump.

\section{Discussion\label{discussion}}

We  analyzed  the high resolution spectra of a sample of 348 stars in terms of their lithium content. 
for which we already had accurate determination of  stellar parameters from  \cite{Zielinski2012}. 
The stars represent a subsample of the 1000 stars being monitored as part of the PTPS program \citep{NW2008}, 
and thus additional information regarding putative companions is available to us. 
The sample of  \cite{Zielinski2012} is composed mainly of regular giant stars evolving along the RGB ($62\pm3\%$)
and  $37\pm3\%$ giants (126 stars) that belong to the RGC according to  extended criteria in luminosity to account 
for uncertainties presented by \cite{Zielinski2012}. It is suitable in searching for any lithium 
enrichment association to a particular evolutionary stage on the RGB.  

After rejecting five dwarfs, 16 stars with uncertain stellar parameters, two SB2 systems and two, subgiants from 
the \cite{Zielinski2012} sample, we performed further  analysis of 323 giants with  spectroscopically determined 
parameters and multiple HET/HRS spectra.

We detect a lithium line in 82 out of the 323 stars analyzed ($25\pm2\%$ of the sample); 11 of these stars are Li-rich according 
to the criteria that $\ALi_{NLTE}\gtrsim1.4$~dex, which represents $13\pm4\%$ of the stars with lithium detection (or $3.4\pm1\% $
of the stars in the full sample). Seven  stars ($2.2\pm1\%$) reveal Li abundances above the commonly 
accepted threshold of $\ALi>1.5$.  We also identify a group of four ($1.4\pm1\%$
of the sample) super-Li giants with $\ALi> 2.0$ (within uncertainties), which share many characteristics, such as 
high mass, luminosity, effective temperature, and metallicity, and are all RV variable.
These giants include TYC 0684-00553-1 and TYC 3105-00152-1, 
stars  with Li abundances close to meteoritic level, and TYC~3304-00090-1 and TYC~3917-01107-1,
giants with $\ALi\approx2$ and planetary systems.

We have located these stars on the HR diagram, correlated their properties in terms of lithium content 
with the stellar parameters in the search for trends, performed accurate abundance analysis 
of $\alpha$ elements, looked for infrared excesses, and finally contextualized the Li-rich sample within 
the additional information provided, since these stars are part of the PTPS program. 

Within the narrower range of stellar effective temperatures for the stars in our sample, we find similar 
behavior in terms of lithium abundances and $\Teff$, as it has been already pointed out in the 
literature \citep{Brown1989, Lebzelter2012, Gonzalez2009}, which is lower in lithium abundances for cooler stars.
 We have divided the sample in three groups: one, containing all the stars where we provide 
 upper limits for Li abundance; the second, composed of those stars with Li detection 
 but $\ALi_{NLTE} \lesssim 1.4$~dex; and a third, containing all the Li-rich giants of the sample. 

Regarding the stellar parameters, we find no substantial deviation in the cumulants 
of different groups in terms of their distribution of luminosity, metallicity, abundances
of Al, O, Mg, Ti, and Ca,  and rotational velocities. 
We find, however, differences in effective temperature, gravitational acceleration, 
microturbulence velocity, stellar mass, and radii.
Giants with detected Li favor higher  effective temperatures and gravitational accelerations
but lower radii.  Li-rich giants tend to have even higher effective temperatures and  larger 
masses compared to giants with no lithium, and they have larger $\vmic$ than giants with detected Li.
It is important to note, however, that for most of the stars  we do not have reliable parallaxes,
and therefore, we have an important source of uncertainty in their masses, radii, and luminosities. 

The most common scenario for lithium production in giants is the Cameron-Fowler mechanism.
It is expected to  operate in particular locations on the HR diagram that are usually defined as 
the  RGC or LFB, where extra-mixing is expected to take place. 
The  LFB occurs on the RGB for star with masses $<2.2\Msun$, which probably has a dependency 
with metallicity, and takes place no earlier than  $\Teff \gtrsim 4400$~K \citep{CharBal2000}. 

Thirty-five ($43\pm5\%$) out of 82 giants with detected Li fall into RGC and only 28 (9$\pm2\%$) into LFB. 
Most of the 11 giants with the largest Li content and luminosities of $\log L/L_{\odot}=1-1.5$ are likely first ascent RGB stars 
that will ignite He core fusion at the tip of RGB.
 Only four of them (36$\pm$15$\%$) fall into RGC and one (9$\pm$9$\%$) into LFB.}
 
Within uncertainties, these fractions agree with the complete sample, and we conclude that 
Li-detected giants are distributed on HRD, as a complete sample with no preference toward RGC or LFB.
The location of the Li-rich stars of our sample,therefore, agrees with the general conclusions 
of  \cite{ Anthony-Twarog2013}, \cite{Martell2013}, and \cite{Lebzelter2012}, who find that Li-rich giants can be found 
almost anywhere along the RGB and that they are not particularly associated to a particular location on the HR diagram.

On the other hand, the  LFB is nothing but the place where an additional mechanism of mixing, which is not defined 
in standard stellar evolution theory, takes place. \cite{Eggleton2008} showed that  the removal of the molecular 
weight discontinuity is responsible for the observed changes in abundances of various elements,
 such as in the decline of the ratio of carbon isotopes and the enhancement in nitrogen abundance.
 According to the recent model of extra mixing driven by fast internal rotation by \cite{Denissenkov2012},
 abundances of aluminum might be also a clue for extra mixing process. However, we do not see 
 any trends in aluminum abundances for Li-rich stars, which may be associated with extra mixing.
 Observed statistical differences in [Al/Fe] distributions for Li-poor, detectable, and rich stars might reflect 
 that both Li moderate and rich stars cluster in narrow [Fe/H] range.

External pollution from the type II supernova as the origin of the Li-enrichment should reveal a correlation 
with the $\alpha$-element abundances. We find no conclusive systematic correlation between the abundances 
of the $\alpha$-elements and the groups established that regards the presence of lithium, although there is the
exception of calcium and titanium. 
If anything, Li-detected stars in our sample seem to show a preference toward lower $\alpha$-element 
abundances; however, according to the Kolmogorov-Smirnov tests, this relation is not very strong. 
Furthermore, oxygen and magnesium seem to be less abundant for stars, where lithium can be measured,
regardless of the lithium abundance level. If that is a consequence of the inhomogeneous stellar populations 
present in our sample, it is difficult to address with the data at hand and the lack of accurate distances for most of the stars.
There is no sign that Li overabundance is connected to supernovae explosions in neither of 
the selected subsamples nor the individual cases. 

Our chemical composition analysis shows  also that the enrichment
from a more evolved stellar companion \citep{SacBoo1999}  mechanism is not  
common  among Li-rich giants. Of the 11 stars with significant Li-overabundance, only one, TYC~3314-01371-1 
is a spectroscopic binary and shows a very modest oxygen overabundance.

We have also explored the possibility that the lithium abundances might be associated to engulfment 
of companions that would have a consequence of a mass loss increase. 
Engulfing low-mass companions is a standard consequence of the increasing radius of the star that
climbs the RGB and of tidal interactions \citep{VillaverLivio2009}.
As this effect would most
likely be shown as an infrared excess, we have to search for them using the 2MASS, WISE, and IRAS catalogs, 
and we do not find any strong evidence in the data of this being the case. None of the 11 Li-rich giants shows K-WISE[12$\mu$m]  
excess. In the de La Reza et al. (1997) diagram only one  of Li-rich stars (TYC~3917-1107-1) is present in the high IR excess region II.
Of the 78 giants with IRAS data available, only ten with detected Li ($16\pm5\%$)  belong to the region II, where moderate IR excess is expected. 
That fraction is lower than in the total sample ($25\pm2\%$)
and we can state that giants with detected Li or overabundance do not show a trace 
of a more intense mass loss than giants with no detected Li in general.

Engulfment has been  
explored to explain the existence of fast-rotating giants \citep{Carlberg2009}, as the orbital 
angular momentum of planet might be transferred to a star, which speeds up its rotation. We have 
identified, however, only one Li--rich fast rotator among Li-rich stars (TYC 1058-02865-1, $v \sin i =6 \kms$) and another three 
with $v \sin i >10 \kms$, which are classified as Li-poor stars. Our statistical analysis suggests that Li-rich,
Li-detectable, and Li-poor stars have identical rotational velocities compared to the rest of stars in sample. 
The differences found in velocities are marginal, they do not exceed the $1 \kms$ level.
We did not confirm the relation of Li content with stellar rotation.

More interesting is the  combination of spectroscopic analysis and radial
velocity measurements obtained within  PTPS.
Four of 11 stars in the most Li--rich sample, including two the super-Li giants, present low level RV variations pointing 
to planetary or brown dwarf-mass companions. The giant TYC~3304-00090-1 has been shown to have a planetary mass companion 
in a very eccentric orbit by \cite{Adamow2012}, who suggest that the high Li content in this giant is due to engulfment 
of a hypothetical second planet. The object TYC~3917-01107-1(Niedzielski in prep.) is another example of Li-rich giant with 
a planetary system. The other two Li-rich giants also display RV variations suggesting planetary/brown dwarf  companions.
In all, our data suggests that the presence of companions is an important factor in the enrichment processes. 
It may stimulate the Li production processes through tidal interactions, 
or by the engulfment of planets or brown dwarf. However, given the lack of evidence of enhanced mass-loss or rotation 
in Li-rich giants, the mechanism of chemical enrichment in the course of engulfment certainly requires rethinking.
In principle the Li overabundance in giants  may be 
connected with the hypothesis by \cite{Ramirez2012} that planets may slow down the Li depletion process in 
dwarfs located in the so-called ``lithium desert'' but how that process affects giants remains unclear.

The other seven Li-rich giants  present a more standard population with one spectroscopic binary, 
one star with low amplitude RV variations, and five apparently single stars. 
The binary frequency in the small sample is $ 29 \pm 17\%$ and agrees with a general binary frequency among 
red giants of 21-30\% \citep{MermilliodMayor1992, GunnGriffin1979}. 
The nature of Li enrichment of those six giants is that they has most likely not completed  FDU dilution yet.  
Statistical analysis pointing to higher masses and effective temperatures for Li-detectable and Li-rich stars
support this conclusion. Massive stars spend less time on MS than solar-masses objects; hence, their 
depletion in this evolutionary stage is not as severe. Observed $\ALi$ abundances for $2.5 - 3\Msun$ MS stars 
show Li abundances up to meteoritic level. Their location at the base or in lower RGB with $\ALi \sim 1.4-1.5$
is then not surprising.

Only one star, TYC~0405-01700-1, seems to present characteristics
attributed to several Li-enrichment processed at the same time. 
It has a stellar companion, seems to be fast rotator, and may be reveling some kind of chromospheric activity. 
It is also located in the LFB region, and it is the only star that shows low IR excess.
It shows O overabundance: its source is probably not accretion of supernova remnants but a transfer 
from a stellar companion.
Although this star is a very difficult case to spectroscopic studies, it is  certainly a very interesting object. 
In his work, \cite{Denissenkov2012} suggests that Li rich stars
reveal characteristics similar to RS CVn stars. It may be indeed the case for  TYC~0405-01700-1.

\begin{acknowledgements}
We thank Dr. Nikolai Piskunov and Dr. Jeff Valenti for making SME available for us.
We thank the HET and IAC resident astronomers and telescope operators for support.
We thank the anonymous referee  for comments that resulted in substantial improvement of this paper.

MA, AN and GN were supported in part by the Polish Ministry of Science and Higher Education
grant N N203 510938 and by the Polish National Science Centre grant no. UMO-2012/07/B/ST9/04415.
MA is also supported by the Polish National Science Centre grant no. UMO-2012/05/N/ST9/03836.
GN is also supported by The Faculty of Physics, Astronomy and Informatics of the Nicolaus Copernicus
University grant no. 1627-A. EV work was supported by the Spanish Ministerio de
Ciencia e Innovacion (MICINN), Plan Nacional de Astronomia y Astrof\'isica, under grant AYA2010-20630
and by the Marie Curie grant FP7-People-RG268111. 
AW was supported by the NASA grant NNX09AB36G. The HET is a joint project of the University of
Texas at Austin, the Pennsylvania State University, Stanford University, Ludwig-
Maximilians-Universit\"at M\"unchen, and Georg-August-Universit\"at G\"ottingen.

The HET is named in honor of its principal benefactors, William P. Hobby and Robert E. Eberly.

The Center for Exoplanets and Habitable Worlds is supported by the Pennsylvania State University,
the Eberly College of Science, and the Pennsylvania Space Grant Consortium. 

This research has made use of the SIMBAD database, operated at CDS
(Strasbourg, France) and NASA's Astrophysics Data System Bibliographic Services.

This research has made use of the NASA/IPAC Infrared Science Archive, which is operated
by the Jet Propulsion Laboratory, California Institute of Technology, under contract with the
National Aeronautics and Space Administration.
\end{acknowledgements}

%\bibliographystyle{aa} % style aa.bst
%\bibliography{literature} % your references Yourfile.bib

% 
\longtab{5}{
\begin{landscape}
\begin{longtable}{l | cccc | ccc |cc |cc| cc | cc | cc | c|c  }
\caption{Lithium abundances for 348 PTPS stars.}\\
\hline
No. TYC  & $v\sin i_{\mathrm{CCF}}$ &$\sigma$& $v\sin i_{\mathrm{SME}}$ & $\sigma$ &$\ALi$ &   $\ALi_{NLTE} $ & $\sigma \ALi$& 
                       [O/H]  & $\sigma$ &
                      [Mg/H] & $\sigma$ & [Al/H] & $\sigma$&[Ca/H] & $\sigma$&[Ti/H] & $\sigma$ &Pop.&\\
\hline
\hline
\endfirsthead
\caption{continued.}\\
\hline\hline
No. TYC  & $v\sin i_{\mathrm{CCF}}$ &$\sigma$& $v\sin i_{\mathrm{SME}}$ & $\sigma$ &$\ALi$ &   $\ALi_{NLTE} $ & $\sigma \ALi$& 
                       [O/H]  & $\sigma$ &
                      [Mg/H] & $\sigma$ & [Al/H] & $\sigma$&[Ca/H] & $\sigma$&[Ti/H] & $\sigma$ &Pop.&\\

\hline
\endhead
\hline
\endfoot
%%%%
0014-00693-1 &    3.1 &   0.9 &  1.7 &  0.7 &       $<$0.07 & $<0.25$ & &       -0.18 & 0.13 &       -0.27 & 0.12 &        -0.29 & 0.1 &       -0.35 & 0.17 &       -0.37 & 0.15 &  d & \\ 
0014-00731-1 &    2.8 &   0.9 &  1.7 &  0.7 &       $<$0.42 & $<0.61$ & &       -0.15 & 0.14 &       -0.17 & 0.12 &       -0.19 & 0.09 &       -0.32 & 0.16 &       -0.31 & 0.15 &  d & \\ 
0014-00752-1 &    2.9 &   0.9 &  2.1 &  0.6 &       $<$0.85 & $<1.06$ & &       -0.02 & 0.18 &       -0.16 & 0.12 &        -0.14 & 0.1 &       -0.33 & 0.16 &       -0.27 & 0.16 &  d & \\ 
0014-00769-1 &    2.9 &   0.9 &  2.0 &  0.5 &       $<$0.31 & $<0.55$ & &       -0.24 & 0.16 &       -0.24 & 0.12 &        -0.27 & 0.1 &       -0.42 & 0.17 &       -0.44 & 0.16 &  d & \\ 
0014-00882-1 &    3.8 &   0.9 &  2.9 &  0.3 &      $<$-0.19 & $<0.11$ & &        0.21 & 0.15 &       -0.02 & 0.12 &        0.22 & 0.13 &       -0.09 & 0.15 &        0.03 & 0.15 &  d & \\ 
0017-00572-1 &    2.8 &   0.9 &  2.0 &  0.5 &      $<$-0.22 & $<0.02$ & &       -0.28 & 0.19 &       -0.19 & 0.12 &       -0.18 & 0.09 &       -0.34 & 0.16 &       -0.26 & 0.16 &  d & \\ 
0017-00668-1 &    2.6 &   0.9 &  1.5 &  0.7 &       $<$0.07 & $<0.26$ & &       -0.09 & 0.15 &       -0.24 & 0.12 &       -0.26 & 0.09 &       -0.39 & 0.16 &       -0.33 & 0.15 &  t & \\ 
0017-00900-1 &    2.8 &   0.9 &  2.0 &  0.5 &     $<$-0.27 & $<-0.03$ & &  -0.07 & {\it 0.18} &  -0.19 & {\it 0.12} &       -0.24 & 0.09 &       -0.58 & 0.15 &        -0.4 & 0.17 &  d & \\
\end{longtable}
\tablefoot{$^a$stars with non-LTE corrections for the lowest available Li value in \cite{Lind2009} grids.  Column (17) contains information on the stellar subsystem, to which given star belongs, according to \cite{Ibukiyama2002} criteria: d -- disk, t -- thick disk, h -- halo. Values of uncertainties marked with italics are mean errors for a given element. Flags in column (20): $b$ --stars with uncertain stellar parameters from \cite{Zielinski2012}, $c$ -- dwarfs identified by \cite{Zielinski2012}, ** -- SB2 stars identified in CCF analysis.}
\end{landscape}
}

\end{document}